\documentclass[runningheads]{llncs}
\usepackage{syntax}
\usepackage{mathpartir}
\usepackage{amsmath}
\usepackage{amssymb}
\usepackage{graphicx}
\usepackage{booktabs}
\usepackage{subcaption}
\usepackage{mathtools}
\usepackage{dsfont}
\usepackage{paralist}
\usepackage{caption}
\usepackage{amscd}
\usepackage{tikz}
\usepackage{xparse}
\usetikzlibrary{matrix}
\usetikzlibrary{fit}
\usetikzlibrary{automata}
\usetikzlibrary{arrows,shapes,backgrounds,positioning,calc,decorations.pathmorphing}
\usepackage{multirow}
\usepackage{dashbox}
\usepackage{amsfonts}
\usepackage[normalem]{ulem}
\usepackage{wrapfig}
\usepackage{appendix}
\usepackage{algpseudocode}
\usepackage[linesnumbered,ruled,noend]{algorithm2e}
\algnewcommand\algorithmicforeach{\textbf{for each}}
\algdef{S}[DOWHILE]{ForEach}[1]{\algorithmicforeach\ #1\ \algorithmicdo}
\algdef{SE}{PreLoop}{EndPreLoop}[1]{\textbf{preloop} \(\mbox{#1}\) \textbf{do}}{\textbf{end}}%

\usepackage{listings}
\usepackage[utf8]{inputenc}

\usepackage[T1]{fontenc}
\usepackage{textcase}
\usepackage{empheq}
\usepackage{tcolorbox}
\usepackage{csquotes}

\usepackage{xspace}

\newcommand{\Nats}{\ensuremath{\mathbb{N}}\xspace}

\newcommand{\eg}{e.g.\xspace}
\newcommand{\ie}{i.e.\xspace}

\newcommand{\secref}[1]{Sec.~\ref{sec:#1}}

\newcommand{\figref}[1]{Fig.~\ref{fig:#1}}
\newcommand{\exref}[1]{Example~\ref{ex:#1}}
\newcommand{\algoref}[1]{Algo.~\ref{algo:#1}}
\newcommand{\thmref}[1]{Theorem~\ref{thm:#1}}

\newcommand{\appref}[1]{App.~\ref{app:#1}}

\newcommand{\tabref}[1]{Table~\ref{table:#1}}
\newcommand{\lineref}[1]{Line~\ref{line:#1}}

\newcommand{\para}[1]{\noindent{\bf \em #1}}

\newcommand{\tool}{\textsc{Venus}\xspace}

\newcommand{\squeezeexample}{\vspace{-5pt}}
\newcommand{\squeezeexampleend}{\vspace{-2pt}}

\newcommand{\n}{n}

\newcommand{\consagree}{agreement\xspace}

\newcommand{\action}{\alpha\xspace}

\newcommand{\lhs}{\mathtt{lhs}}
\newcommand{\rhs}{\mathtt{rhs}}
\newcommand{\M}{\mathcal{M}}

\newcommand{\sysname}{QuickSilver}
\newcommand{\kinarach}{\textsc{\sysname}\xspace}

\newcommand{\winset}{W}	

\newcommand{\encompassing}{encompassing\xspace}

\newcommand{\pairwise}{rendezvous\xspace}

\newcommand{\ValueCons}{\terma{Consensus}\xspace}
\newcommand{\valueCons}{\terma{consensus}\xspace}
\newcommand{\PartitionCons}{\terma{Partition}\xspace}
\newcommand{\partitionCons}{\terma{partition}\xspace}

\definecolor{Gray}{gray}{0.85}
\definecolor{LightRed}{rgb}{1,0.7,0.7}
\definecolor{LightGreen}{rgb}{0.7,1,0.7}

\definecolor{realGreen}{rgb}{0.0, 0.5, 0.0}

\newcommand{\mnx}[1]{}

\newcommand{\aml}{\textsc{Mercury}\xspace}

\newcounter{sarrow}

\makeatletter
\newcommand*\bigcdot{\mathpalette\bigcdot@{.5}}
\newcommand*\bigcdot@[2]{\mathbin{\vcenter{\hbox{\scalebox{#2}{$\m@th#1\bullet$}}}}}
\makeatother

\newcommand{\reduceStep}[3]{#1 \xrightarrow{#2} #3}

\newcommand{\termb}[1]{\code{#1}}
\newcommand{\terma}[1]{\textbf{\code{#1}}}

\newcommand{\decVal}{decVar}

\newcommand{\ourskip}{\smallskip}
\newcommand{\fixedskip}{\vspace{2pt}}

\newcommand{\ats}{\mathcal{A}}

\newcommand{\nodes}{\mathcal{L}}
\newcommand{\edges}{\mathcal{E}}
\newcommand{\edge}{e}
\newcommand{\vars}{\mathcal{V}}

\newcommand{\locof}{loc}

\newcommand{\edgeset}{edges}

\newcommand{\region}{r}
\newcommand{\regions}{regions}

\newcommand{\varSet}{\mathcal{V}}

\newcommand{\domSet}{\mathrm{\Delta}}
\newcommand{\dom}{\delta}
\newcommand{\varDom}[1]{vdom(#1)}

\newcommand{\eventWithPayload}[2]{#1[#2]}
\newcommand{\val}{val}

\newcommand{\localState}{s}

\newcommand{\localStateSet}{S}
\newcommand{\localStates}{\localStateSet}
\newcommand{\localInitialStateSet}{\localStateSet_0}
\newcommand{\localInitialStates}{\localInitialStateSet}
\newcommand{\localTrans}{t}
\newcommand{\localTransRel}{T}
\newcommand{\globalState}{q}
\newcommand{\altGlobalState}{r}

\newcommand{\boundedGlobalState}{\overline{\globalState}}
\newcommand{\boundedAltGlobalState}{\overline{\altGlobalState}}
\newcommand{\boundedEvent}{\overline{\event}}

\newcommand{\globalStateSet}{Q}
\newcommand{\globalStates}{\globalStateSet}

\newcommand{\globalTransRel}{R}

\newcommand{\perm}{\pi}

\newcommand{\localPerm}{\perm}
\newcommand{\localPermSet}{G}

\newcommand{\globalPerm}{\gamma}
\newcommand{\cwpSet}{\Gamma}

\newcommand{\boundedRegion}{\Psi}

\newcommand{\projection}{\Pi}
\newcommand{\consistent}{cons}
\newcommand{\cwpRel}{\approx_\boundedRegion}
\newcommand{\shellRel}{\approx}

\newcommand{\regionBound}{\rho}
\newcommand{\localBound}{\lambda}

\newcommand{\program}{P}
\newcommand{\environment}{E}

\newcommand{\directionSym}{\text{\rotatebox[origin=c]{90}{$\multimap$}}}

\newcommand{\direction}[1]{#1\directionSym}

\newcommand{\actingSym}{\text{!}}
\newcommand{\acting}[1]{#1\actingSym}
\newcommand{\reactingSym}{\text{?}}
\newcommand{\reacting}[1]{#1\reactingSym}
\newcommand{\bounded}[1]{\overline{#1}}

\newcommand{\boundedDomSet}{\bounded{\domSet}}

\newcommand{\overbar}[1]{\mkern 1.5mu\overline{\mkern-1.5mu#1\mkern-1.5mu}\mkern 1.5mu}

\newcommand{\sboundedDomSet}{\overbar{\domSet}}

\newcommand{\mv}{\M_{\domSet}}
\newcommand{\mvb}{\M_{\sboundedDomSet}}
\newcommand{\mnv}{\mv(n)}
\newcommand{\mnvb}{\M_{\sboundedDomSet}(n)}
\newcommand{\mnvp}{\M_{\domSet'}(n)}

\newcommand{\Mfamily}{\M(\Nats)}
\newcommand{\boundedMfamily}{\M_{\sboundedDomSet}(\Nats)}
\newcommand{\unboundedMfamily}{\M_{\domSet}(\Nats)}

\newcommand{\pv}{\program_{\domSet}}
\newcommand{\pvb}{\program_{\sboundedDomSet}}

\newcommand{\pd}{\program_{\dom}}
\newcommand{\bpd}{\program_{\bounded{\dom}}}

\newcommand{\varName}{v}
\newcommand{\varDef}{(\varName, \varDom)}
\newcommand{\varDefSet}{\mathcal{V}}
\newcommand{\locVar}{v_{loc}}
\newcommand{\eventSig}{\varepsilon}
\newcommand{\eventSet}{\mathcal{E}}

\newcommand{\event}{e}

\newcommand{\actionSig}[1]{\direction{\eventSig}}

\newcommand{\handler}{h}
\newcommand{\handlerSet}{\mathcal{H}}
\newcommand{\globalSpec}{\Phi}

\newcommand{\unitType}{\terma{unit}\xspace}

\newcommand{\coordType}{\tau}
\newcommand{\actionName}{a}
\newcommand{\eventName}{\code{eID}}
\newcommand{\envFlag}{f}

\newcommand{\intCoordType}{{\tt in}\xspace}
\newcommand{\pwCoordType}{{\tt pw}\xspace}
\newcommand{\bcCoordType}{{\tt bc}\xspace}
\newcommand{\pcCoordType}{{\tt pc}\xspace}
\newcommand{\vcCoordType}{{\tt vc}\xspace}

\newcommand{\localGuard}{\phi}

\newcommand{\updates}{U}

\newcommand{\participantSet}{Prt}
\newcommand{\winSet}{W}

\newcommand{\domainOrder}{\Upsilon_{\dom}}

\newcommand{\process}{P}

\newcommand{\getsFrom}{gets}

\newcommand{\dst}{dst}

\newcommand{\minimal}{min}

\newcommand{\abr}{ABR\xspace}
\newcommand{\abrs}{ABRs\xspace}

\newcommand{\pairs}[1]{#1}

\newcommand{\serverRegion}{serverRegion}

\newcommand{\atsnode}{x}

\newcommand{\evSigTuple}[4]{(#1,#2,#3,#4)}

\newcommand{\dab}{DAB\xspace}
\newcommand{\reducible}{domain-reducible\xspace}
\newcommand{\mkperm}{mk\gamma}

\definecolor{bluekeywords}{rgb}{0.13,0.13,1}
\definecolor{greencomments}{rgb}{0,0.5,0}
\definecolor{turqusnumbers}{rgb}{0.17,0.57,0.69}
\definecolor{redstrings}{rgb}{0.5,0,0}

\definecolor{qualifiers}{rgb}{0.5,0,0.63}
\definecolor{types}{rgb}{0.5,0,0}
\definecolor{communication}{rgb}{0.5,0,0}
\definecolor{categories}{rgb}{0.5,0,0.63}
\definecolor{controlstmts}{rgb}{0.5,0,0}
\definecolor{handlermeta}{rgb}{0.13,0.13,1}
\definecolor{setops}{rgb}{0.5,0,0}
\definecolor{cons}{rgb}{0.5,0,0}

\definecolor{hlcolor}{rgb}{0.13,0.13,1}

\lstdefinelanguage{FSharp}
    {keywords=[1]{env, initial}, 
    keywordstyle=[1]\bfseries\color{qualifiers},
    keywords=[2]{int, idset, id, bool, unit},
    keywordstyle=[2]\bfseries\color{types}, 
    keywords=[3]{sendrz, sendbr,recv,br,rz}, 
    keywordstyle=[3]\bfseries\color{communication},
    keywords=[4]{machine, variables, actions,events, location, satisfies,process},
    keywordstyle=[4]\bfseries\color{categories},
    keywords=[5]{if,else,goto},
    keywordstyle=[5]\bfseries\color{controlstmts},
    keywords=[6]{on, win, lose, passive, do, where},
    keywordstyle=[6]\bfseries\color{handlermeta},
    keywords=[7]{add, remove, con=tains,decVar,winS,loseS,payld, default},
    keywordstyle=[7]\bfseries\color{setops},
    keywords=[8]{partition, consensus},
    keywordstyle=[8]\bfseries\color{cons},
    sensitive=false,
    morecomment=[l][\color{greencomments}]{///},
    morecomment=[l][\color{greencomments}]{//},
    morestring=[b]",
    stringstyle=\color{redstrings}
    }

    \lstnewenvironment{dsl}
  {
    \lstset{
        language=FSharp,
        basicstyle=\ttfamily,
         numbers=left,
    	stepnumber=1,
        breaklines=true,
        escapeinside={{(*}{*)}},
        columns=fullflexible
        }
        }
  {
  }

    \lstnewenvironment{udsl}
  {
    \lstset{
        language=FSharp,
        basicstyle=\ttfamily,
        breaklines=true,
        escapeinside={{(*}{*)}},
        columns=fullflexible
        }
        }
  {
  }

    \lstnewenvironment{sdsl}
  {
    \lstset{
        language=FSharp,
        basicstyle=\ttfamily\footnotesize,
         numbers=left,
    	stepnumber=1,
        breaklines=true,
        escapeinside={{(*}{*)}},
        columns=fullflexible
        }
        }
  {
  }

      \lstnewenvironment{ssdsl}
  {
    \lstset{
        language=FSharp,
	 basicstyle=\ttfamily\scriptsize,
         numbers=left,
    	stepnumber=1,
        breaklines=true,
        escapeinside={{(*}{*)}},
        columns=fullflexible
        }
        }
  {
  }

     \lstnewenvironment{sudsl}
  {
    \lstset{
        language=FSharp,
        basicstyle=\ttfamily\tiny,
        breaklines=true,
        escapeinside={{(*}{*)}},
        columns=fullflexible
        }
        }
  {
  }

\definecolor{blue-violet}{rgb}{0.54, 0.17, 0.89}

\newcommand{\code}[1]{\texttt{#1}} 
\newcommand{\reducespace}{-1pt}
\newcommand{\squeezecaption}{\vspace{\reducespace}}
\newcommand{\reallyreducespace}{-10pt}
\newcommand{\reallysqueezecaption}{\vspace{\reallyreducespace}}

\begin{document}

\title{Bounded Verification of Doubly-Unbounded Distributed Agreement-Based Systems}
\titlerunning{Bounded Verification of Doubly-Unbounded \dab Systems}



\author{Christopher Wagner\and
	Nouraldin Jaber \and
	Roopsha Samanta
}
\authorrunning{C. Wagner, N. Jaber, and R. Samanta}

\institute{Purdue University, West Lafayette, USA\\ \email{\{wagne279,njaber,roopsha\}@purdue.edu}}

\maketitle

\begin{abstract}

The ubiquity of distributed agreement protocols, such as consensus,
has galvanized interest in verification of 
such protocols {\em as well as} applications built on top of them. 
The complexity and unboundedness of such systems, however, 
makes their verification onerous in general, and, particularly prohibitive for full automation.
An exciting, recent breakthrough reveals that, through careful modeling, 
it becomes possible for verification of interesting distributed agreement-based (\dab) systems, 
that are unbounded in the number of processes, 
to be reduced to model checking of small, finite-state systems.

It is an open question if such reductions are also possible for 
\dab systems that are {\em doubly-unbounded}, in particular, 
\dab systems that additionally have unbounded data domains. 
We answer this question in the affirmative in this work for 
models of \dab systems, 
thereby broadening the class of \dab systems 
which can be automatically verified. 
We present a new symmetry-based reduction and
develop a tool, \tool, that can efficiently verify 
sophisticated \dab system models.

\end{abstract}

\section{Introduction}\label{sec:intro}


A recent breakthrough in formal reasoning about distributed systems builds on the modularity inherent in their design to enable {\em modularity in their verification}~\cite{Jaber.QuickSilver.2021,TLC.Griffin.ICFP.2020,Sergey.ProgrammingProvingDistributed.POPL.2017}.
The central approach incorporates {\em abstractions} of common core protocols, such as distributed consensus, into verification of applications {\em built on top of} such core protocols. 
This approach inspires an interesting epiphany: modular models of distributed systems based on protocol abstractions may permit {\em fully-automated verification \`a la model checking}, even when their monolithic and {\em intricate} counterparts do not~\cite{GSP,Jaber.QuickSilver.2021}.
Amenability to full automation is significant because the problem of algorithmically verifying correctness of systems with an unbounded number of processes, popularly known as the parameterized model checking problem (PMCP), is a well-known undecidable problem \cite{Suzuki.PMCP.UndecidableFirstPPR.1988.PPR}. 
For instance, recent work~\cite{Jaber.QuickSilver.2021} identifies a class of modular models of distributed services, based on {\em distributed agreement protocols}, for which PMCP can be reduced to model checking of small, finite-state systems. 
However, a limitation of this approach (as well as most decision procedures for PMCP \cite{EsparzaFM99,FS01,GSP})
is that it requires each process in the distributed system model to be finite-state. This stipulation is not surprising---the distributed system becomes {\em doubly-unbounded} if, in addition to an unbounded number of processes, it has infinite-state processes.

In this paper, we seek to substantially expand the class of unbounded distributed agreement-based (\dab) systems which can be automatically and scalably verified through {\em modular and bounded verification}.
Towards this goal, we characterize modular models of \dab systems, unbounded in the number of processes {\em as well as} variable domains whose verification can be reduced to that of small, bounded systems.

We tackle PMCP for systems that are unbounded along two dimensions, one dimension at time. 
We first focus on the verification of \dab systems with unbounded variable domains and some fixed number $n$ of processes.
%
We identify conditions under which their verification can be reduced to model checking of $\n$-process systems with small, bounded variable domains.
Intuitively, our approach has the following features.

\fixedskip
\para{Value Symmetry}.
We first analyze if a system exhibits {\em symmetry} in its use and access of variables with unbounded domains, specifically, if the system's correctness is preserved under 
permutations of values from these domains. As is standard, verification of such symmetric systems can exploit the induced redundancies, for instance, through the use of a smaller quotient structure.

\fixedskip
\para{Data Saturation}. 
A quotient structure is not guaranteed to be finite-state in general.
Hence, we seek to check if our value-symmetric system also exhibits {\em data saturation}; that is, if the size of its symmetry-reduced quotient structure has a finite upper bound, even when the systems’ data domains are unbounded.

\fixedskip
\para{Domain Cutoffs}.
Finally, to enhance the applicability of our {\em domain reduction}, we automatically infer bounds on the data domains of the system's processes. We refer to these bounds as {\em domain cutoffs}. We argue that domain cutoffs
are more {\em flexible} than an upper bound for a symmetry-reduced quotient structure. 

With the data domains reduced to a fixed, finite space, it now becomes possible to leverage existing results~\cite{Jaber.QuickSilver.2021} for PMCP of \dab systems with an unbounded number of finite-state processes. Specifically, PMCP can now use \emph{process cutoffs} to reduce verification to that of systems with a fixed, finite number of processes. We emphasize that our domain reduction and cutoffs are not restricted to be used only with \cite{Jaber.QuickSilver.2021}; instead, our inference of domain cutoffs can facilitate flexible combination with varied techniques to obtain process cutoffs. 

To summarize, we make the following contributions for \dab system models: 
\begin{compactenum}
\item Symmetry-based Domain Reduction (~\secref{value-cutoffs}): We characterize systems with unbounded data domains that exhibit {\em symmetry} and {\em saturation} in their data usage, thereby enabling a reduction of their verification to that of systems with finite, bounded domains.
\item Domain Cutoffs (~\secref{value-bounding}): We present a sound procedure to infer bounds on the data domains of systems that permit our symmetry reduction. 
\item \tool (~\secref{eval}): We develop a tool, \tool, for parameterized verification of doubly-unbounded systems that can efficiently compute domain/process cutoffs and verify sophisticated \dab systems. 
\end{compactenum}

\section{Illustrative Overview}
\label{sec:examples}

\begin{figure*}[ht]
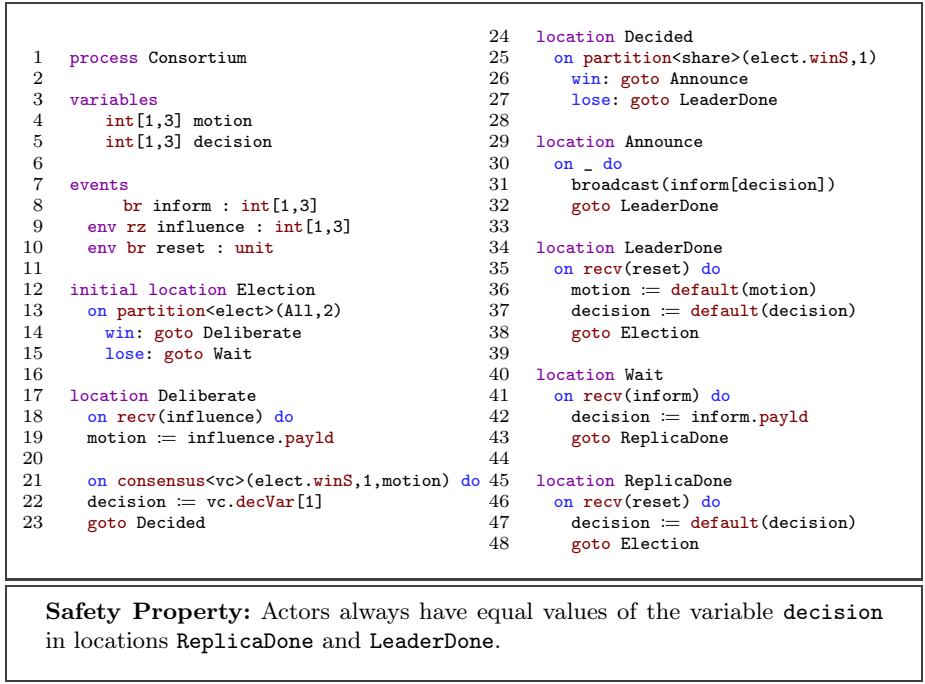

\centering
\begin{tcolorbox}[colback=white,sharp corners,boxrule=0.3mm,top=-.1mm,bottom=-.1mm]
\hspace{1em}\begin{minipage}{0.55\textwidth}
\begin{ssdsl}
process Consortium

variables 
    int[1,3] motion
    int[1,3] decision
    
events  
      br inform : int[1,3]  (*\label{line:informDecl}*)
  env rz influence : int[1,3]
  env br reset : unit 	

initial location Election (*\label{line:election}*)
  on partition<elect>(All,2)  (*\label{line:elect}*)
    win: goto Deliberate
    lose: goto Wait   (*\label{line:gotoWait}*)
		
location Deliberate (*\label{line:deliberate}*)
  on recv(influence) do (*\label{line:influence}*)
  motion (*$\coloneqq$*) influence.payld

  on consensus<vc>(elect.winS,1,motion) do(*\label{line:vc}*)
  decision (*$\coloneqq$*) vc.decVar[1]
  goto Decided
\end{ssdsl}
\end{minipage}
\begin{minipage}{0.70\textwidth}
 \lstset{firstnumber=24}
\begin{ssdsl}
location Decided (*\label{line:decided}*)
  on partition<share>(elect.winS,1) (*\label{line:share}*)
    win: goto Announce
    lose: goto LeaderDone
    
location Announce    
  on _ do
    broadcast(inform[decision]) (*\label{line:inform-send}*)
    goto LeaderDone
    
location LeaderDone (*\label{line:leaderdone}*)
  on recv(reset) do (*\label{line:leader-reset}*)
    motion (*$\coloneqq$*) default(motion)
    decision (*$\coloneqq$*) default(decision)
    goto Election
  
location Wait (*\label{line:wait}*)
  on recv(inform) do (*\label{line:inform-recv}*)
    decision (*$\coloneqq$*) inform.payld
    goto ReplicaDone
    
location ReplicaDone (*\label{line:replicadone}*)
  on recv(reset) do (*\label{line:replica-reset}*)
    decision (*$\coloneqq$*) default(decision)
    goto Election
\end{ssdsl}
\end{minipage}
\end{tcolorbox}

\squeezeexample

\begin{tcolorbox}[colback=white,sharp corners,boxrule=0.3mm, top=1mm, bottom = 2.5mm]
\footnotesize{
	\textbf{Safety Property:}
	Actors always have equal values of the variable \termb{decision} in locations \termb{ReplicaDone} and \termb{LeaderDone}.
}
\end{tcolorbox}

\squeezecaption
\squeezecaption

\caption{\aml model of a Consortium process.
	}
\label{fig:consortiumcode} 
\squeezecaption

\end{figure*}

%

%
%

\noindent We motivate and illustrate our contributions with an example distributed system, {\em Consortium}, that uses different types of distributed agreement to achieve trust-based consensus.

\subsection{Motivation: Distributed Agreement-Based Systems}

\para{Consortium}. The Consortium distributed system involves a set of actors who want to mutually make a decision based on information they gather individually. In order to do this efficiently, a subset of the actors is elected and trusted with making the decision and announcing it to the rest of the actors. This resembles scenarios where a trade-off between trust and performance is needed (e.g., a consortium blockchain~\cite{libraref,hyperledgerref}). 

We model each actor/process of Consortium in \aml~\cite{Jaber.QuickSilver.2021}, a modeling language with inbuilt primitives for distributed agreement. As can be seen in \figref{consortiumcode}, \aml facilitates a clean, {\em modular} design of \dab systems such as Consortium with its encapsulation of the intricacies of agreement protocols into agreement primitives.

An actor initially starts in the \code{Election} location and coordinates with all other actors (\lineref{elect}) to elect at most two
actors. Notice that this election is performed using a ``\partitionCons'' agreement primitive with identifier \code{elect} (\lineref{elect}); this instance of \partitionCons expresses that \code{2} actors are elected from among \code{All} actors. 
The elected actors move to the \code{Deliberate} location where they are trusted to make a decision for everyone. 
The remaining actors move to the \code{Wait} location (\lineref{gotoWait}), where they wait to be informed once the elected actors agree on a decision (\lineref{inform-recv}). 
The elected actors in \code{Deliberate} may be influenced (by the environment) to update their proposal, stored in variable \code{motion} (\lineref{influence}). Further, the elected actors make a decision using a ``\valueCons'' agreement primitive with identifier \code{vc} (\lineref{vc}); this instance of \valueCons models agreement on \code{1} value proposed in the \code{motion} variable of processes elected in \code{Election}.
After storing the decided value in the \code{decision} variable, the elected actors move to \code{Decided} where they elect one actor (\lineref{share}) to announce the agreed-upon value to all actors (\lineref{inform-send}). Next, the elected actors move to \code{LeaderDone} and all other actors move to \code{ReplicaDone}. All actors then go back to the initial location, reinitializing their variables, to start further rounds (Lines \ref{line:leader-reset}, \ref{line:replica-reset}).

\fixedskip
\para{Correctness Specification and Parameterized Verification}. 
The correctness specification of interest for a Consortium distributed system $\M(n)$ with $n$ instantiations of the above actor is a safety property: all $n$ actors 
in locations \code{ReplicaDone} and \termb{LeaderDone} agree on the value of the variable \termb{decision}. 

We wish to ensure that a Consortium distributed system with an arbitrary, {\em unbounded number of actors} is correct. In particular, we are interested in {\em fully-automated parameterized verification} of Consortium that seeks to algorithmically check if $\M(n)$ satisfies its correctness specification for all values of $n$. Furthermore, we are interested in {\em modular verification} that effectively exploits the modularity of \aml's Consortium model. \fixedskip

\para{Prior Work: Unbounded Number of Processes and Bounded Domains}.
While there exist multiple algorithms and tools for parameterized verification, we are aware of only two fully-automated, modular approaches~\cite{GSP,Jaber.QuickSilver.2021} 
that can tackle \dab systems, of which only one (\kinarach~\cite{Jaber.QuickSilver.2021}) can tackle systems modeled in \aml. In fact, \kinarach can perform parameterized verification of Consortium, as modeled in~\figref{consortiumcode}, {\em efficiently}, in less than a second. 

Sadly, \kinarach is limited to \aml systems composed of {\em finite-state processes} with a relatively modest number of states.
Thus, even though \kinarach can do verification of systems  parameterized with an unbounded number of processes, it cannot handle systems with an unbounded number of processes {\em and} variable domains.  Notice that a Consortium actor is finite-state with integer datatypes over small subranges. If the integer subrange datatypes of variables \code{motion} and \code{decision} in~\figref{consortiumcode} are replaced with true integers (or even standard 32-bit integers), \kinarach is unable to perform parameterized verification of the new system. 

\fixedskip
\para{This work: Unbounded Number of Processes and Unbounded Domains.}
The inspiration for this work is two fold: (a) the effectiveness of \aml in enabling modular design and verification of \dab systems and (b) the confirmation provided by \kinarach that {\em fully-automated parameterized verification of such systems is possible!}
We ask a natural question to help address \kinarach's limitations: 
\begin{displayquote} 

{\em Is fully-automated, modular, and scalable parameterized verification possible for \dab 
systems with large, unbounded, or infinite-state processes?
}

\end{displayquote}
In particular, can we algorithmically and efficiently perform parameterized verification of a {\em doubly-unbounded} version of Consortium where the the integer subrange datatypes of variables and events are replaced 
with (unbounded) integers? In what follows, we continue to denote the Consortium distributed system with $n$ instantiations of the finite-state actor in~\figref{consortiumcode} as $\M(n)$. Further, we denote Consortium with $n$ instantiations of the corresponding infinite-state actor (with unbounded integer datatypes) as $\mnv$.

\subsection{Contributions}

 
\para{Domain Reduction}.
Our key contribution employs notions of {\em value symmetry} and {\em data saturation}, to enable a {\em reachability-preserving transformation} of \aml systems with unbounded variable domains to \aml systems with finite domains. We illustrate these concepts and our {\em domain reduction} on our target (doubly-unbounded) Consortium system.

Notice that an actor of $\mnv$ uses and accesses integer datatypes in a restricted way, in particular, as a {\em scalarset datatype}~\cite{ip1996better}---expressions over this datatype are restricted to (dis)equality checks and variables. 
Notice also that the safety specification accesses integer-valued variables as scalarsets. 
This ``value symmetry'' in an actor of $\mnv$ as well as the safety specification implies preservation of correctness under permutations of scalarsets---an execution $\eta$ of $\mnv$ violates the safety specification iff any execution corresponding to a permutation of the scalarset values in $\eta$ violates the safety specification.

This preservation of reachability under scalarset permutations suggests that many system executions are redundant for verification, and not all of them need to be explored. Quotient structures are used commonly to exploit such redundancies to verify properties of symmetric systems more efficiently. 
For value-symmetric systems, if the number of {\em distinct} values appearing in any reachable state has a {\em finite upper bound}, the size of the quotient structure can have a finite upper bound, even when the systems' scalarset domains become unbounded or infinite; this property is called \emph{data saturation} \cite{ip1996better}. In particular, if such an upper bound can be established on the number of distinct scalarset values appearing in the reachable states of $\mnv$, then safety verification of $\mnv$ can be reduced to safety verification of a system with finite integer domains! 


We now argue that a bound of \emph{three}
distinct values can be established for $\mnv$, thereby reducing safety verification of $\mnv$ to safety verification of $\M(n)$ with integer subrange \code{[1,3]}.
Any execution of $\mnv$ is equivalent (w.r.t. the safety property) to one in which only three distinct integer values appear. This
stems from the  fact that
(a) only \emph{one} value at a time is needed to represent the most recent result of consensus, and (b) each actor has only \emph{two} local variables: \code{motion} and \code{decision}. While, in general, different actors may have distinct values in their local variables, because these values are never compared or communicated between actors (except during consensus), any execution in which actors have more than two distinct values in their local variables can be shown to be equivalent to an execution in which the actors have the same two values in their local variables. Ergo, to verify the unbounded-domain system $\mnv$, it suffices to the verify the [\code{1,3}]-bounded-domain system $\M(n)$.

In this work, we identify conditions under which value symmetry and data saturation enable domain reduction for safety verification of \aml systems with unbounded or infinite-state processes. 
In particular, if (a) all values from an unbounded scalarset domain held by a process in the system can be partitioned into two {\em regions} at any point in a system execution---a region of {\em globally known} values of which {\em all} processes are aware and a region of {\em locally known} values which may be known only to a single process---and (b) the maximum sizes of these two regions can be statically bounded, then the ``large'' unbounded scalarset domain can be replaced by a ``small'' finite one with size equal to the sum of the bounds of these two regions.
We then show that if an unsafe execution exists in the ``large'' system, there must exist a related unsafe execution in the ``small'' system. 

\fixedskip
\para{Domain Cutoffs}.
To support practical application of domain reduction, we present a procedure which analyzes \aml processes to identify two regions as described above and their
associated bounds. 
In doing so, we provide a concrete path to verification of \aml models of \dab systems. As explained above, the two bounded regions yield \aml systems with finite, bounded variable domains, which can then be verified using existing parameterized verification engines such as \kinarach.

\fixedskip
\para{\tool and Evaluation}.
We have implemented a tool, \tool, that combines our domain reduction and cutoffs with \kinarach and can perform fully-automated, modular, and scalable parameterized verification of \aml system models with large, unbounded, or infinite-state processes. In particular, \tool is able to verify, in under a second, that $\mnv$ satisfies its correctness specification for all values of $n$. 

\section{Problem Definition}
\label{sec:mercury}

%

\aml is a modeling language for \dab systems consisting of an arbitrary number of identical processes and built on top of verified \consagree protocols.
 As a native feature, \aml includes two special event handlers, \partitionCons and \valueCons, which capture the general behavior of common \consagree protocols (e.g., leader election and consensus).
In what follows, we first review the syntax and semantics of \aml systems, and then define the parameterized verification problem for such systems.


\subsection{\aml Systems}
\label{sec:mercury-programs}
A \aml process definition $\program$ 
has three main components: (i) a set of typed process-local variable declarations, (ii) a set of typed {\em event signatures} defining the set of coordination events which may occur during program execution, and (iii) a set of {\em action handlers} describing the behavior of a process when an event is initiated or received. 
\fixedskip

\para{Variable Declarations.}\label{sec:var-decls}
Each variable declaration consists of a variable name $\varName$ and an associated variable domain 
which may be either a bounded integer range or the powerset of the set of process identifiers (i.e. $\varName$ may store a set of process identifiers). We denote the set of variables as $\varDefSet$.

\fixedskip
\para{Event Declarations and Actions.}
\label{sec:events}
Event declarations correspond to four different types of process coordination: 
%
(i) pairwise communication, denoted \pwCoordType, where one process sends a message, and one other process receives it, 
(ii) broadcast communication, denoted \bcCoordType, where a process sends a message, and all other processes receive it,
(iii) \partitionCons coordination, denoted \pcCoordType, where a set of participating processes designates a finite subset of themselves as ``winners'', and all other participating processes as ``losers'', or 
(iv) \valueCons coordination, denoted \vcCoordType, where each process in a set of participating processes proposes a value, and all participating processes agree on a finite subset of the proposed values. 
%
%

Each event declaration consists of an event type (i.e., one of \pwCoordType, \bcCoordType, \pcCoordType, \vcCoordType), event name $\eventName$, a payload data type 
(either a bounded integer range or $\unitType$), and optionally a 
keyword \code{env} if the event is an interaction with the environment. Each event declaration describes a set of events and a set of {\em actions}. 
An event $\event \coloneqq \eventWithPayload{\eventName}{\val}$ corresponds to a particular payload value $\val$ of the associated payload data type. 

An action is a {\em polarized} event where the polarity indicates if the event is acting (denoted $\acting{\event}$) or reacting (denoted $\reacting{\event}$) to the event. In particular, acting (resp. reacting) events are sends (resp. receives) of broadcasts and pairwise communication, and correspond to ``winners'' (resp. ``losers'') of agreement coordination\footnote{A process winning consensus means that it proposed one of the agreed-upon values.}.

We denote the sets of events, acting events, reacting events, and actions as $\eventSet$, $\acting{\eventSet}$, 
$\reacting{\eventSet}$, and $\direction{\eventSet}$.
\begin{example}
\squeezeexample
Consider the \code{inform} event declaration defined on \lineref{informDecl} in \figref{consortiumcode}. The event has type $\bcCoordType$, name \code{inform}, and payload type \code{int}. 
The set $\eventSet$ of events associated with this event declaration (and induced by the values of the payload) is $\{\eventWithPayload{\code{inform}}{0},\eventWithPayload{\code{inform}}{1},\ldots\}$. For each event $\event \in \eventSet$, there is an acting and a reacting action. For instance, the acting and reacting actions for event $\eventWithPayload{\code{inform}}{0}$ are  $\acting{\eventWithPayload{\code{inform}}{0}}$ and  $\reacting{\eventWithPayload{\code{inform}}{0}}$, respectively.
\squeezeexampleend
\end{example}
\fixedskip

\para{Action Handlers.}
\label{sec:handlers}
Each location in the process definition is associated with a set of action handlers. 
An action handler comprises an action name, a guard, and a set of updates. A guard is a Boolean predicate over variables in $\varSet$ and the set of updates is essentially a {\em parallel assignment} of expressions to each variable.\fixedskip

\para{\aml Semantics.} 
We refer to the semantics of a \aml process as {\em local} and the semantics of a \aml system composed of multiple identical \aml processes as {\em global}. 

The local semantics of a \aml process, $\program$ is given by a labeled state-transition system $M_\program = (\localStates, \localInitialStates, \localTransRel)$ with a state space $\localStates$, a set $\localInitialStates$ of initial states, and a set $\localTransRel \subseteq \localStates \times \direction{\eventSet} \times \localStates$ of transitions induced by the set of action handlers. 
The state space $\localStates$ corresponds to all possible valuations of variables in $\varDefSet \, \cup \, \{\locVar\}$, where $\locVar$ is a special variable defined to store the location. We denote the value of a variable $\varName$ in a local state $\localState \in \localStates$ as $\localState(\varName)$. 
Action handlers for an event $\event$ induce acting and reacting transitions in $\localTransRel$ labeled with $\acting{\event}$ and $\reacting{\event}$, respectively. In particular, $\localTransRel$ contains a transition $(\localState, \direction{\event}, \localState')$ for $ \direction{\event} \in \{\acting{\event}, \reacting{\event}\}$ iff there exists an action handler for $\direction{\event}$ such that the handler's guard is $true$ in $\localState$ and $\localState'$ is obtained by applying the handler's updates to $\localState$.


The global semantics of a \aml system consisting of $n$ identical processes $\program_1, \ldots, \program_n$ and an environment process $\environment$ is given by a labeled transition system $\M(n) = \langle \globalStateSet, \globalStateSet_0, \globalTransRel \rangle$ describing their parallel composition. Here, 
$\globalStateSet = \localStateSet^n \times \localStateSet_\environment$ is the set of global states, $\globalStateSet_0 = \localStateSet_0^n \times \localStateSet_{0, \environment}$ is the set of initial global states, and $\globalTransRel \subseteq \globalStates \times \eventSet \times \globalStates$ is the global transition relation capturing the process coordination necessary for different event types and payloads.
For instance, 
\begin{compactenum}[1.]
%
\item  $\globalTransRel$~contains a ``broadcast transition'' $(\globalState, \event, \globalState')$ for broadcast event $\event$ iff (1) one process $\program_i$ has a local ``broadcast send'' transition $(\globalState[i], \acting{\event},\globalState'[i])$ and (2) all other processes $\program_j$ have corresponding local ``broadcast receive'' transitions $(\globalState[j], \reacting{\event},\globalState'[j])$ with $\globalState'[j]$ updated with the payload value. 

\item $\globalTransRel$~contains a ``\ValueCons transition'' $(\globalState, \eventName[V'], \globalState')$ for \ValueCons event $\event = \eventWithPayload{\eventName}{V'}$, participant set $\participantSet$, and set of winning values $V'$ iff 
(1) each participating process in $\participantSet$ has a consistent view of the other participants, 
(2) every value $v$ in $V'$ is proposed by some process in $\participantSet$ with a local transition for $\acting{\event}$,
(3) each process in $\participantSet$ has a corresponding local $\reacting{\event}$ local transition, and 
(4) the local states of all other processes remain unchanged. 

\end{compactenum}

\begin{figure}[t]
\centering
\includegraphics[width=0.85\linewidth]{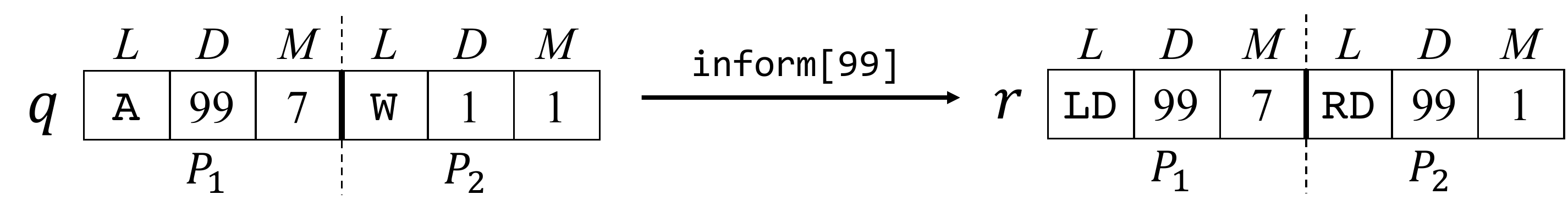}
\caption{A broadcast transition for the event $\event = \code{inform}[99]$ . Here, L, D, and M refer to the $\locVar$, \code{decision}, and \code{motion} variables resp., and \code{A}, \code{W}, \code{LD}, and \code{RD} refer to the \code{Announce}, \code{Wait}, \code{LeaderDone}, and \code{ReplicaDone} locations, resp.}
\label{fig:transitionExFig}
\squeezecaption
\squeezecaption
\reallysqueezecaption

\end{figure}

\begin{example}
\squeezeexample
Consider a Consortium system with two processes $\process_1$ and $\process_2$. \figref{transitionExFig} shows a broadcast transition on event $\event = \code{inform}[99]$. Since $\process_1$ has local transition  $(\globalState[1], \acting{\event}, \altGlobalState[1]) \in \localTransRel$ and 
$\process_2$ has local transition  $(\globalState[2], \reacting{\event}, \altGlobalState[2]) \in \localTransRel$, the global transition $(\globalState, \event, \altGlobalState)$ is in the global transition relation $\globalTransRel$.
\end{example}

We refer the reader to~\appref{semanticsApp} for the complete local and global \aml semantics. An execution of a global transition system $\M(n)$ is defined in a standard way. A global execution is a (possibly infinite) sequence of global states, $\globalState_0, \globalState_1, \ldots$ in $\globalStateSet$ such that for each $j \geq 0$, ($\globalState_j,\event,\globalState_{j+1}) \in \globalTransRel$ for some event $\event$. Global state $\globalState$ is reachable if there exists a finite execution of $\M(n)$ that ends in $\globalState$.



\subsection{The Parameterized Verification Problem for \aml Systems}
For a system $\M(n)$ with some number $n \in \Nats$ of {\em finite-state} processes $\program$ and a correctness specification $\globalSpec$, we use  $\M(n) \models \globalSpec$ to denote that the system $\M(n)$ satisfies $\globalSpec$. The {\em parameterized model checking problem} (PMCP) 
targets the verification of a family $\Mfamily$ of systems 
$\{\M(0),\M(1),\ldots\}$ w.r.t. correctness specification $\globalSpec$. In particular, PMCP seeks to check if $\forall n.\, \M(n) \models \globalSpec$~\cite{BloemETAL15}. Note that this standard formulation of PMCP assumes that each process $\program$ has a finite-state space. In order to enable reasoning about \aml processes with unbounded and possibly infinite state spaces, we introduce new notation and a new formulation for the parameterized verification problem. 

%
%
%

We denote a \aml process with a set $\domSet$ of (possibly infinite) data domains as $\pv$ and a process with a set $\boundedDomSet$ of finite data domains as $\pvb$. We denote a \aml system with $n$ instances of processes $\pv$ (resp. $\pvb$) as $\mnv$ (resp. $\mnvb$). In this paper, we target the parameterized verification problem over a family $\unboundedMfamily$ of \aml systems 
$\{\mv(0),\mv(1),\ldots\}$, defined as: 
$$
  \forall n. \; \mnv \models \globalSpec.
$$

The above problem generalizes PMCP for \aml systems from the verification of an infinite family $\boundedMfamily$ of finite-state systems $\{\mvb(0),\mvb(1),\ldots\}$ to an infinite family $\unboundedMfamily$ of infinite-state systems $\{\mv(0),\mv(1),\ldots\}$.

%


\section{Domain Cutoffs for \aml Systems}
\label{sec:value-cutoffs}

To enable parameterized verification over a family $\unboundedMfamily$ of \aml systems, we utilize value symmetry and data saturation to present a reduction of verification of the infinite-state \aml system $\mnv = \langle \globalStateSet, \globalStateSet_0, \globalTransRel \rangle$ 
to verification of a finite-domain \aml system. In particular, we characterize verification problems, denoted $\langle \mnv,\Phi \rangle$, that are {\em \reducible} and  hence permit a reachability-preserving transformation that replaces the data domains of $\mnv$ with finite, bounded ones. 
Let $\mnvb = \langle \bounded{\globalStateSet}, \bounded{\globalStateSet}_0, \bounded{\globalTransRel} \rangle$ denote the \aml system with $n$ finite-state processes $\pvb$, where $\pvb$ is obtained from $\pv$ by replacing the latter's data domains $\domSet$ with finite data domains 
$\boundedDomSet$. We show the following.



\begin{theorem}
	\label{thm:safetyThm}
  For \reducible $\langle \mnv,\Phi \rangle$, 
	$\mnv \models \globalSpec \Leftrightarrow \mnvb \models \globalSpec$
\end{theorem}
In what follows, we present a proof of this claim.
Our proof relies on establishing a {\em backward simulation} relation $\shellRel$ 
over pairs of global states in $\mnv$ and $\mnvb$:
For all $\altGlobalState \in \globalStateSet, \boundedAltGlobalState \in \bounded{\globalStateSet}$, if $\altGlobalState \shellRel \boundedAltGlobalState $ and $(\globalState, \event, \altGlobalState) \in \globalTransRel$, then there is 
$(\boundedGlobalState, \boundedEvent, \boundedAltGlobalState) \in \bounded{\globalTransRel}$
such that
$\globalState \shellRel \boundedGlobalState$.
This relation enables a proof of \thmref{safetyThm} by induction on an arbitrary error path in $\mnv$ to show that a related error path exists in $\mnvb$. 

To simplify presentation, we assume, initially, that $\domSet$ contains exactly one unbounded, scalarset domain $\dom$. Further, we elide the environment process, which is not impacted by our reachability-preserving transformation.

\subsection{Permutations and Scalarsets}

We first review standard notions of permutations and scalarsets. Then, we introduce a notion of {\em component-wise} permutations.
All notions of permutations are w.r.t. {\em the} unbounded, symmetric domain $\dom \in \domSet$.\fixedskip

\para{Permutation.} 
A $\dom$-permutation, $\localPerm: \dom \mapsto \dom$, is a bijection mapping the set $\dom$ onto itself.
With some abuse of notation, we set $\localPerm(\val) = \val$ for any value $\val$ with domain in $\domSet \setminus \dom$. 
We further define liftings of $\dom$-permutations to local states and events. An application $\localPerm(\localState)$ of a $\dom$-permutation $\localPerm$ to a local state $\localState$ is defined as: $\forall \varName \in \varDefSet:
\localPerm(\localState)(\varName) = \localPerm(\localState(\varName))$. 
An application $\localPerm(\event)$ of a $\dom$-permutation $\localPerm$ to an event $\event = \eventName[\val]$ is $\eventName[ \localPerm(\val)]$, the permutation $\localPerm(\event\directionSym)$ of an action of event $\event$ is $\localPerm(\event)\directionSym$, and the permutation of a local transition $(\localState, \direction{\event}, \localState')\in  \localTransRel$ is $(\localPerm(\localState), \localPerm(\direction{\event}), \localPerm(\localState'))$.
\begin{wrapfigure}[11]{r}{0.17\textwidth}
\squeezeexample
\squeezeexampleend
\squeezeexampleend
\squeezeexampleend
\squeezeexample
\centering
\includegraphics[width=\linewidth,clip, trim=1cm 2.1cm 2.1cm 1.3cm]{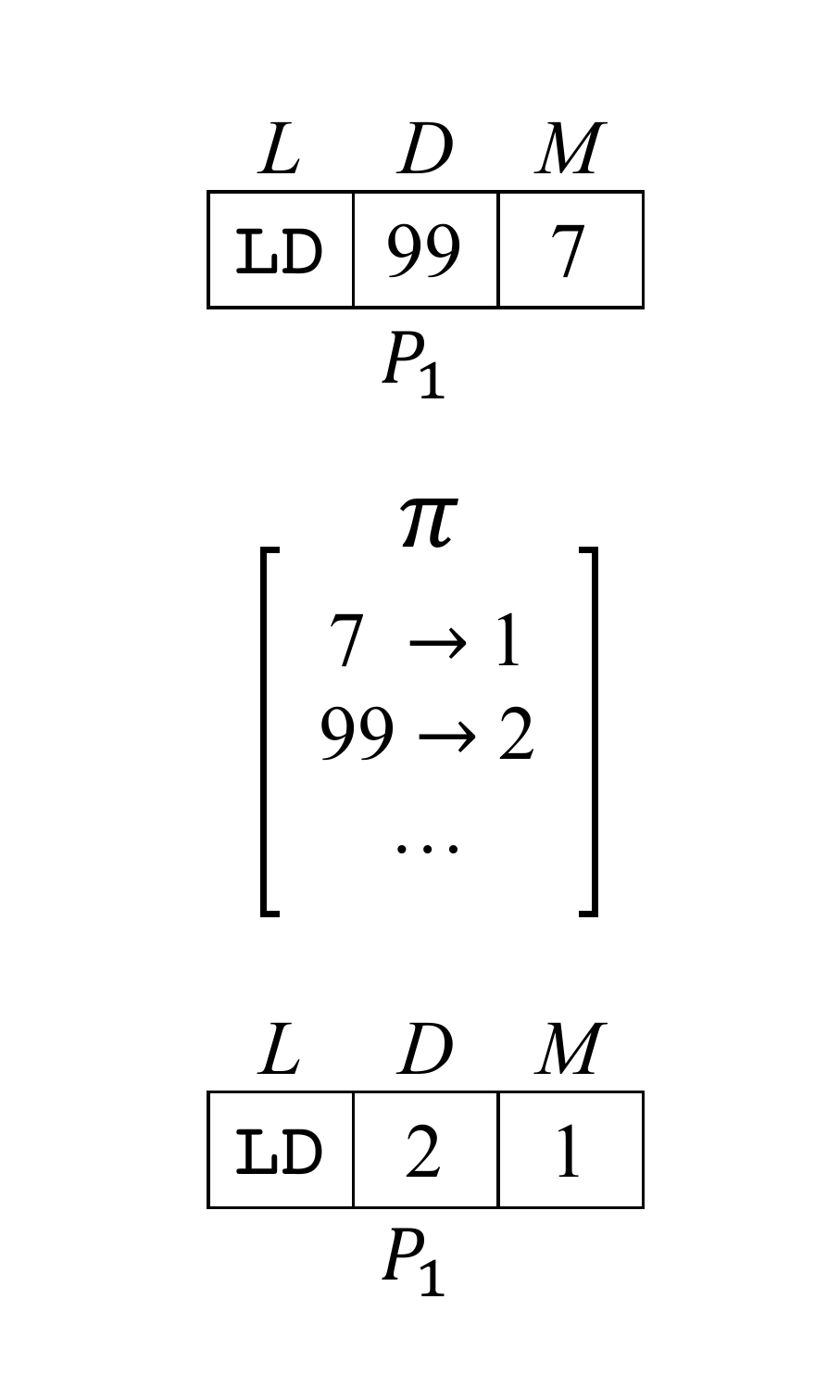}
\label{fig:localPermFig}
\end{wrapfigure}
\squeezeexample
\squeezeexampleend
\squeezeexample
\begin{example}
We illustrate permutations on our running example, Consortium. The figure to the right shows an example of applying a $\dom$-permutation $\localPerm$ to the local state of process $\process_1$ from \figref{transitionExFig}. Let $\dom$ denote the type \code{int}. Recall that the local variables \code{decision} and \code{motion} (denoted $D$ and $M$ in the figure) are of type \code{int}.
The permutation $\localPerm$ maps values 7 to 1, 99 to 2, and all other values appropriately so that $\localPerm$ is a valid permutation (i.e., a bijection over the \code{int} domain). Notice that the value of the variable $\locVar$ (denoted $L$ in the figure) is not of type \code{int} and, hence, it is not changed in the permuted state (i.e. $\localPerm(\text{\code{LD}}) = \text{\code{LD}}$).
 \squeezeexampleend
\end{example}

\para{Scalarsets.}
A scalarset domain \cite{ip1996better} is a set of distinct elements with restricted operations. Specifically, (i) all valid scalarset terms are variable references; there are no scalarset constants,
(ii) scalarset terms may only be compared using (dis)equality and only with terms of
the same scalarset type, and (iii) scalarset variables may only be assigned values of exactly the same scalarset type.
These restrictions ensure that the local transition relation is invariant over permutations of a scalarset:
\begin{lemma}
	\label{lem:local-trans-preserved}
		$\forall \localPerm \in \localPermSet, (\localState,\ \action,\ \localState') \in  \localTransRel:
		\localPerm((\localState,\ \action,\ \localState')) \in \localTransRel
		$
\end{lemma}
For the proof, see \appref{proofs-trans-local}. We note that the bounded integer domains in \aml can be treated as scalarsets if used according to the restrictions above. 
For instance, the \code{decision} variable in \figref{consortiumcode} is of type \code{int} but can be treated as a scalarset variable because it conforms to the constraints (i) through (iii). 
\ourskip

\begin{wrapfigure}[14]{r}{0.32\textwidth}
	\centering
	
	\squeezeexample
	\squeezeexample
		\includegraphics[width=\linewidth,clip, trim=0.3cm 0cm 0cm 0cm]{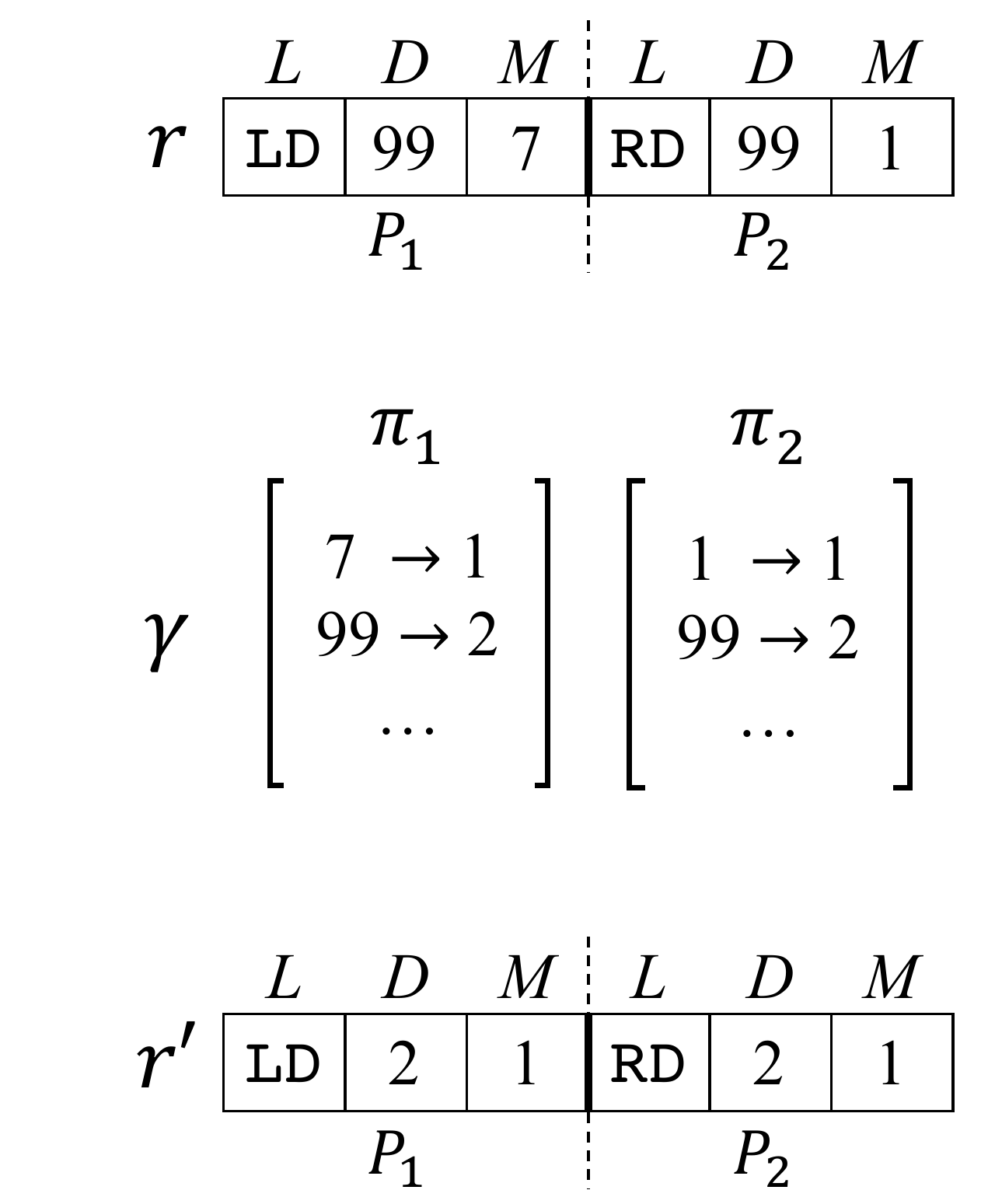}
	\caption{A CWP.}
	\label{fig:globalPermFig}
\end{wrapfigure}

\para{Component-wise Permutation.}
Next, we define a new type of transformation on global system states, called a {\em component-wise permutation} (CWP). CWPs 
consist of a series of separate $\dom$-permutations 
which are applied, component-wise, to each local state in a global state. For a global state $\globalState = (\localState_1, \localState_2, \ldots, \localState_n)$   
let $\globalPerm = (\localPerm_1, \localPerm_2, \ldots, \localPerm_n)$ be a CWP over the values of $\dom$ and $\globalPerm(\globalState)$ be the component-wise application of 
permutations in $\globalPerm$ to local states in $\globalState$. That is, $\globalPerm(\globalState) = (\localPerm_1(\localState_1), \localPerm_2(\localState_2), \ldots, \localPerm_n(\localState_n))$.
Informally, one can think of a CWP as a tool permute a global state $\globalState$ in $\mnv$ with an unbounded data domain to a global state $\globalPerm(\globalState)$ in the target system, $\mnvb$, with a bounded data domain, effectively ``collapsing'' it. Such a reduction from unbounded to bounded domains is possible because a CWP is allowed to permute different values in different components to the same value, as shown in the following example. \phantom{don't remove this.}
\begin{example}
\label{ex:cwp-example}

An example CWP $\globalPerm = (\localPerm_1, \localPerm_2)$ applied to a global state $\altGlobalState$ of Consortium, yielding $\altGlobalState'$, is shown
in \figref{globalPermFig}. 
The state of process $\process_1$ (resp. $\process_2$) is permuted according to $\localPerm_1$ (resp. $\localPerm_2$). Observe that the larger values in $\altGlobalState$ are permuted to smaller values in $\altGlobalState'$. Further, notice that while there are three distinct values in $\altGlobalState$, there are only two in $\altGlobalState'$!
\end{example}

\subsection{Scalarset Domain Reduction}
\label{sec:vsr}
%
%

While, in general, a CWP $\globalPerm$ can permute the same value in different components to different permuted values, 
at certain global states in an execution, equality between some values in different processes is significant (e.g., when the results of consensus are determined). Hence, component permutations of $\globalPerm$ must permute such values \emph{consistently}
to preserve their equality
(i.e., if such values are equal before applying $\globalPerm$, they should be after $\globalPerm$ is applied).
Next, we characterize these \emph{regions} in the execution where values must be permuted consistently.
\fixedskip

\para{Value-Stable and Bounded Regions}.
A \emph{region} $\boundedRegion \subseteq \localStateSet \times \varSet$ is a set of pairs of process-local states and variables.
Given a global state $\globalState$, we denote the set of $\dom$-values appearing in $\globalState$ as $\projection_\dom(\globalState)$, and appearing in the local state  $\globalState[i]$ 
 of the $i^{th}$ process as 
$\projection_\dom(\globalState[i])$. 
 Given a global state $\globalState$ and a region $\boundedRegion$, 
we denote the set of $\dom$-values appearing in corresponding elements of $\boundedRegion$ as $\projection^\boundedRegion_\dom(\globalState)$, 
\ie, $\projection^\boundedRegion_\dom(\globalState)$ $=$ $\{\val = \localState(\varName) \mid (\localState, \varName) \in \boundedRegion \text{\,and\,} \exists i: \globalState[i] = \localState\}$. 
%
%

We say a region $\boundedRegion$ is \emph{value-stable} if there exists a finite upper bound $\regionBound$ on the number of distinct values in $\projection^\boundedRegion_\dom(\globalState)$ over all 
reachable global states $\globalState$.
Additionally, a value-stable region $\boundedRegion$ is \emph{bounded}
if there exists a finite upper bound, $\localBound$, 
over all reachable global states $\globalState$, 
on the number of distinct $\dom$-values that are 
held by any single process and are 
\emph{not} in $\projection^\boundedRegion_\dom(\globalState)$. 
More precisely, for a bounded, value-stable region, there exists $\localBound$ such that for all reachable global states $\globalState$: 
$\forall i: |\projection_\dom(\globalState[i]) \setminus \projection^\boundedRegion_\dom(\globalState)| \;  \leq  \; \localBound$. 
\fixedskip

\para{Consistent and Minimal CWPs.}
For a given region $\boundedRegion$,
we say that a CWP
$\globalPerm$
is \emph{consistent} w.r.t. a global state $\globalState$, denoted $\consistent(\globalPerm, \globalState, \boundedRegion)$, iff all component permutations of $\globalPerm$ map each value in 
$\projection_\dom^{\boundedRegion}(\globalState)$
to the {\em same} permuted value. 
%
Formally, 
$
\consistent(\globalPerm, \globalState, \boundedRegion)$ iff  
$\forall i, j, 
\forall \val \in \projection_\dom^{\boundedRegion}(\globalState):
\globalPerm[i](\val) = \globalPerm[j](\val).
$
Additionally, we say that a CWP $\globalPerm$
is \emph{minimal} with respect to a global state $\globalState$, 
denoted $\minimal(\globalPerm,\globalState, \boundedRegion)$, 
iff each component permutation $\localPerm_i$ of $\globalPerm$ 
permutes all values in $\projection_\dom(\globalState[i]) \setminus 
 \projection_\dom^\boundedRegion(\globalState)$
to the smallest {\em available} values 
\ie, to the smallest values, according to an arbitrary total order over $\dom$, 
that are not in 
the image of any component permutation. 
%
%
%
%
\begin{example}
\squeezeexampleend

 For \figref{globalPermFig}, consider a bounded region $\boundedRegion$ 
 that contains the state-variable pair $(\altGlobalState[2],\code{decision})$. A CWP $\globalPerm$ is consistent w.r.t $\altGlobalState$ iff all the component permutations of $\globalPerm$ permute the value 99 (in the \code{decision} variable in $\altGlobalState[2]$) consistently to the same permuted value (e.g., to 2 as done by $\globalPerm = (\localPerm_1, \localPerm_2)$). On the other hand, if $\localPerm_1$ is changed to permute 99 to, say 3 instead of 2, then the resulting $\globalPerm$ is not consistent. 
	Further, assume that $\boundedRegion$ {\em does not contain} the state-variable pairs $(\altGlobalState[1],\code{motion})$ and $(\altGlobalState[2],\code{motion})$. The CWP $\globalPerm$ in \figref{globalPermFig} is minimal since it permutes the values outside of $\boundedRegion$ (7 and 1 in 
	$(\altGlobalState[1],\code{mention})$ and $(\altGlobalState[2],\code{mention})$, respectively) to the smallest available value, 1.
	If
	$\localPerm_1$ is changed to permute 7 to, say 3 instead of 1,
	then the resulting $\globalPerm$ would not be minimal, because $\localPerm_1$ could have permuted 7 to a smaller value, 1.
	\squeezeexampleend
\end{example}

\para{Consistency.}
We denote by $\cwpSet_\dom$ the set of all CWPs over the domain $\dom$.
For a given region $\boundedRegion$,
we define the corresponding consistency set  $\cwpSet_\dom^\boundedRegion(\globalState)$ 
w.r.t. a global state $\globalState$ 
to be the set of CWPs
which are consistent and minimal w.r.t $\globalState$. 
That is, 
$\cwpSet_\dom^\boundedRegion(\globalState) = \{\globalPerm
\in \cwpSet_\dom \mid \consistent(\globalPerm, \globalState, \boundedRegion) \land \minimal(\globalPerm,\globalState, \boundedRegion)\}$.

For a domain $\dom$ and bounded region $\boundedRegion$, we define an equivalence relation $\cwpRel$,
over pairs of global states in $\mnv$ and $\mnvb$, derived from the consistency sets of global states in $\mnv$: 
$
\cwpRel = \{(\globalState, \boundedGlobalState)  \in \globalStateSet \times \bounded{\globalStateSet} \mid \exists \globalPerm \in \cwpSet_\dom^\boundedRegion(\globalState):  \globalPerm(\globalState) = \boundedGlobalState\}.
$

\fixedskip

\para{Scalarset Domain Reduction.}
For the system $\mnv$ composed of $n$ instances of $\pd$, 
we say that a region $\boundedRegion$ is \emph{\encompassing} if it is value-stable and bounded,  and
it satisfies a set of conditions (detailed in section 5.1) which ensure
its consistency set preserves the global semantics of $\mnv$.
In other words, for any $\globalState \in \globalStates$ and CWP $\globalPerm \in \cwpSet_\dom^\boundedRegion(\globalState)$, $\globalPerm$ consistently permutes any values which may have been received in $\globalState$, and if $\globalState$ is an initial (resp. error) state, $\globalPerm(\globalState)$ is also an initial (resp. error) state.

Then,
 let us \emph{assume} the existence of
an
\encompassing
region $\boundedRegion$,
with bounds $\regionBound$ and $\localBound$ as defined earlier. When such a bounded region $\boundedRegion$ exists for $\langle \mnv, \globalSpec \rangle$, we say that $\langle \mnv, \globalSpec \rangle$ is \emph{\reducible}.
Let $\bounded{\dom}$ be a scalarset domain of size $\regionBound + \localBound$ and. 
We denote by $\bpd$, the process obtained from $\pd$ by replacing all variable and event declarations of type $\dom$  with declarations of type $\bounded{\dom}$. Let $\mnvb$ be the system composed of $n$ instances of $\bpd$. 
%
%
We then show that safety verification of $\mnv$ can be reduced to verification of $\mnvb$.

\begin{lemma}
\label{lem:backward-congruence}

$
\forall \altGlobalState \in \globalStateSet, \boundedAltGlobalState \in \bounded{\globalStateSet}, 
(\globalState, \event, \altGlobalState) \in \globalTransRel:
\altGlobalState \cwpRel \boundedAltGlobalState
\implies
(\exists (\boundedGlobalState, \boundedEvent, \boundedAltGlobalState) \in \bounded{\globalTransRel}:
\globalState \cwpRel \boundedGlobalState).
$
\end{lemma}


\noindent{\em Proof sketch.} 
Since $\altGlobalState \cwpRel \boundedAltGlobalState$, we know there must exist a CWP $\globalPerm$ such that $\globalPerm(\altGlobalState) = \boundedAltGlobalState$.
Then, we must identify an appropriate global state $\boundedGlobalState$ and $\boundedEvent$ such that $(\boundedGlobalState, \boundedEvent, \boundedAltGlobalState) \in \bounded{\globalTransRel}$ and $\globalState \cwpRel \boundedGlobalState$. 
Since $\boundedGlobalState$ must be related to $\globalState$ by $\cwpRel$, we need to show that there exists another CWP $\globalPerm' \in \cwpSet_\dom^\boundedRegion(\globalState)$ such that 
$\globalPerm'(\globalState) = \boundedGlobalState$. 
By identifying $\globalPerm'$, $\boundedEvent$ such that $(\globalPerm'(\globalState), \boundedEvent, \boundedAltGlobalState) \in \bounded{\globalTransRel}$, the lemma is proven by letting $\boundedGlobalState = \globalPerm'(\globalState)$.
We carefully define $\globalPerm'$ to (i) agree with $\globalPerm$ on values preserved in the transition from $\globalState$ to $\altGlobalState$ so that $\globalPerm'(\globalState)$ may transition to $\boundedAltGlobalState$, (ii) permute values in $\projection_\dom^\boundedRegion(\globalState)$ consistently 
and 
values outside of $\projection_\dom^\boundedRegion(\globalState)$ to the smallest possible values so that 
$\globalPerm' \in \cwpSet_\dom^\boundedRegion(\globalState)$.
With $\globalPerm'$ in hand, it remains to show that there exists an event $\boundedEvent$ such that the transition $(\globalPerm'(\globalState), \boundedEvent, \boundedAltGlobalState) $ is a valid transition in $ \bounded{\globalTransRel}$, which can be achieved by permuting each transmitted value to match the permutation (in $\globalPerm$) of a process that receives it. We refer the reader to \appref{backward-congruence} for the full proof.

\fixedskip
\para{Reducing Multiple Domains.}
We focused on the single reduction 
$\reduceStep{\mnv}{\dom}{\mnvb}$ with respect to unbounded scalarset domain $\dom \in \domSet$.
Our results extend immediately to the case with multiple unbounded data domains via a sequence of reductions w.r.t. each $\dom \in \domSet$:
$\reduceStep{\mnv}{\dom_1}{\reduceStep{\mnvp}{\dom_2}{\reduceStep{\ldots}{\dom_m}{\mnvb}}}.$

\section{Determining Bounded Regions}

\label{sec:value-bounding}
%
%
%
Recall that if there exists a bounded region $\boundedRegion$ with (i) an upper bound $\regionBound$, over all reachable global states $\globalState$, on the number of unique $\dom$-values $\projection^\boundedRegion_\dom(\globalState)$ stored in variables of processes in $\boundedRegion$,
 and (ii) an upper bound $\localBound$, over every local state $\localState$ in any reachable global state $\globalState$, on the number of unique $\dom$-values \emph{not} in $\projection^\boundedRegion_\dom(\globalState)$, 
then we can reduce the \reducible verification problem $\langle \mnv, \globalSpec \rangle$ to the verification for the bounded-domain system $\mnvb$,
where $|\bounded{\dom}| = \regionBound + \localBound$, w.r.t. $\globalSpec$.  In this section, we discuss how such a bounded region $\boundedRegion$ and bounds $\regionBound$, $\localBound$ may be determined. \fixedskip

\para{Location Transition System (LTS).}
To determine $\boundedRegion$ with bound $\regionBound$, we inspect the movement of values in the \aml system. 
In particular, we wish to examine values that may {\em flow} 
into a set of local states, \emph{without} constructing the unbounded local semantics of $\pv$. 
So, we begin by defining a finite-state transition system $\ats$ capturing the {\em control-flow} of a \aml process: 
$\ats = \langle \nodes, \edges \rangle$, 
where each node in $\nodes$ corresponds to a location in $\pv$ and
there exists an edge $(a,b)$ in $\edges$ iff 
there exists a transition in the local semantics of $\pd$ 
between states in locations $a$ and $b$.
Let the location of node $\atsnode \in \nodes$ be denoted $\locof(\atsnode)$ and 
the set of local states with $\locVar = \locof(\atsnode)$ be denoted $[\atsnode]$. 
To track relevant data flow, we extend $\ats$ with some bookkeeping information for each edge, in particular, the set of action handlers yielding the edge. 
Let the set of edges derived from a handler $h$ be $\edgeset(h)$.
Notice that $\ats$ constitutes an over-approximation of the local semantics of $\pv$\footnote{The LTS constructed in our implementation is refined further based on the local guards in the process definition. We omit detailing this refinement to simplify the presentation.}.




\begin{example}
\squeezeexampleend
The \code{Partition} handler (\figref{consortiumcode}, \lineref{share}) yields an edge $\edge = (n_1, n_2) \in \edges$, where $\locof(n_1) = \code{Decided}$ and $\locof(n_2) = \code{Announce}$.
\label{ex:locStruc}
\squeezeexampleend
\end{example} 

Finally, we extend our notion of a region from \secref{vsr} to LTSs. Let an \emph{abstract region} $\region \subseteq \nodes \times \vars$ be a set of node-variable pairs such that each node-variable pair $(\atsnode,\varName)$ represents the set  $\{(\localState,\varName) \mid \localState \in [\atsnode] \}$ of state-variable pairs. Let $[\region]$ be the concrete region $\{(\localState,\varName) \mid (\localState,\varName) \in [(\atsnode,\varName)] \wedge (\atsnode, \varName) \in \region \}$ that the abstract region $\region$ represents. We say an abstract region $\region$ is value-stable (resp. bounded) iff the corresponding concrete region $[\region]$ is value-stable (resp. bounded). The bounds $\regionBound(\region)$ and $\localBound(\region)$ of an abstract region $\region$ are equal to the bounds $\regionBound$ and $\localBound$ of $[\region]$.



\subsection{Domain-Cutoff Analysis}

%
%
%
%
%
%


We now present a procedure for determining a suitable  abstract bounded region (\abrs), defined in \algoref{valBounding}.
%
\setlength{\textfloatsep}{4pt}
\begin{algorithm}[t]
\small
  \SetKwFunction{mapproc}{Rewrite}
  \SetKwProg{myproc}{procedure}{}{}
  \myproc{getAbstractBoundedRegion($\ats,\pd$)}{
  
  \SetKwInOut{Inputs}{Inputs}
  \SetKwInOut{Output}{Output}
  \Inputs{A process $\pd$, and the LTS $\ats$}
  \Output{$\region$, an \abr}
  \BlankLine

  
  
  $\regions = $ \Call{getInitialRegions}{$\pd, \ats$}
  
  $\regions = $ \Call{expandRegions}{$\regions, \ats$}
    
  $\regions = $ \Call{mergeRegions}{$\regions, \ats$}
  
  $\region = $ \Call{getMinimalRegion}{$\regions$}  

  \Return $\region$;

}

\caption{Determining an abstract bounded region $\region$}
\label{algo:valBounding}

\end{algorithm}
\fixedskip

\para{Initial Regions.} We identify a set of (initial) abstract regions where for each region $\region$ in that set, it is clear that \emph{any} number of processes in $\region$ have a finite number of distinct values of type $\dom$ among the associated variables of $\region$ (i.e., $\regionBound(\region)$ is finite).
To that end, we identify three types of such initial \abrs as defined in  
\algoref{init-regions}. 
First, for each \valueCons action $vc$, create an \abr composed of all the nodes corresponding to locations where $vc$ terminates along with the variable holding the decided values (\lineref{init-reg-cons}). The bound $\regionBound$ of this \abr matches the cardinality of the \valueCons action $vc$. Intuitively, by the nature of \valueCons, we know that these variables hold the agreed-upon values, which are consistent across participating processes in these locations.
Second, if the locations corresponding to a set of nodes can only be occupied by a single process at a time (e.g. a server), we create an \abr for each of these nodes with every variable of type $\dom$ in the system (\lineref{init-reg-server}). The bound $\regionBound$ of each of these \abrs is the number of variables of type $\dom$, as one process can only hold that many unique $\dom$-values.
%
Finally, we create a single \abr (with $\regionBound=1$) encoding variables of type $\dom$ in the initial location, where all processes hold some initial default value (\lineref{init-reg-static}).\fixedskip

\para{Expanding Regions.} Initial regions are expanded to maintain the bound $\regionBound$. Intuitively, we follow the flow of values between variables so that any node-variable pair that \emph{only} gets values from some \abr region is added to that region (\lineref{expand-getsfrom}). We denote by $\getsFrom(\atsnode, \varName, \atsnode', \varName')$ that for some $\localState' \in [\atsnode']$ the value $\localState'(\varName')$ goes to variable $\varName$ in some state $\localState \in [\atsnode]$ via a direct assignment or a transmission.

\fixedskip
\para{Merging Regions.} If two \abrs $\region_1$ and $ \region_2$ are mutually exclusive (i.e., processes are only in one of them at a time), we create a new \abr $\region = \region_1 \cup \region_2$ with $\regionBound(\region) = max(\regionBound(\region_1),\regionBound(\region_2))$. Since processes may not be in $\region_1$ and $\region_2$ simultaneously, their bounds 
are independent and the larger  applies to both regions.

\begin{algorithm}[t]
\small
  \SetKwRepeat{Do}{do}{while}
  \SetKwProg{myproc}{procedure}{}{}
  \myproc{getInitialRegions($\pd,\ats$)}{
  
  \SetKwInOut{Inputs}{Inputs}
  \SetKwInOut{Output}{Output}
  \Inputs{A \aml program $\pd$ and an LTS $\ats$}
  \Output{a set of \abrs}
  \BlankLine

    $\regions = \{\}$
    
    \ForEach{$a \in consActions(\pd)$}{ \label{line:init-reg-cons}
    
        $H = handlersOf(a)$
        
        $\pairs{\region} = \{(\dst(\edge),a.\decVal)\mid \handler \in H \land \edge \in \edgeset(\handler)\}$              
        
        $\regionBound(\region) = cardOf(h)$

        $\regions = \regions \cup \region$          
                
    }   

    \ForEach{$\localState \in \serverRegion(\ats)$}{ \label{line:init-reg-server}

        $\pairs{\region} = \{\localState\} \times \vars$
    
        $\regionBound(\region) = |\vars|$
    
        $\regions = \regions \cup \region$              
            
    }

    $\pairs{\region} = \{(\atsnode,\varName)\ \mid \not \exists (\atsnode',\varName') : \getsFrom(\atsnode,\varName,\atsnode',\varName') \} $ \label{line:init-reg-static}

    $\regionBound(\region) = 1$

    $\regions = \regions \cup \region$          
    
    \Return $\regions$
}
\caption{Constructing initial regions}
\label{algo:init-regions}
\end{algorithm}

\begin{algorithm}[t]
\small
  \SetKwRepeat{Do}{do}{while}
  \SetKwProg{myproc}{procedure}{}{}
  \myproc{expandRegions($\regions, \ats$)}{
  
  \SetKwInOut{Inputs}{Inputs}
  \SetKwInOut{Output}{Output}
      \Inputs{The LTS $\ats$ and an initial set $\regions$ of regions}
  \Output{a set of \abrs}
  \BlankLine

    \ForEach{$\region \in \regions$}{
    
        \Do{$changeMade$}{
            
            $changeMade = false$            
            
            \ForEach{$(\atsnode,\varName) \not \in  \pairs{\region}$}{
            
                
                \If{$\{(\atsnode',\varName')\mid \getsFrom(\atsnode,\varName,\atsnode',\varName') \} \subseteq \pairs{\region}$}{ \label{line:expand-getsfrom}
                
                    $\pairs{\region}= \pairs{\region} \cup \{(\atsnode,\varName)\}$

                    $changeMade = true$         
                
                }   
                
            }   
            
        }
        
    }

    \Return $\regions$
}
\caption{Expanding regions}
\label{algo:exp-regions}
\end{algorithm}

\fixedskip

\para{A Minimal Region.} After constructing a set of \abrs, we select those which meet a few conditions to support the domain-reduction technique from \secref{value-cutoffs}. In particular, for some \abr $\region$, we require that the corresponding bounded region $[\region]$
must include (i) all initial local states, (ii) all destination local states of consensus actions, (iii) all source local states of broadcast actions with payloads, and (iv) all state-variable pairs referred to by the global specification $\globalSpec$. We select the final \abr $\region$ that satisfies these criteria and has the minimal bound $\regionBound$ of all such candidate \abrs. The bounded region $\boundedRegion$ is then $[\region]$ with bound $\regionBound = \regionBound(\region)$. 
We determine the local bound $\localBound$ by simply counting the number of variables of type $\dom$ in the program $\pd$ that appear in state-variable pairs outside of $\boundedRegion$.
\ourskip

\begin{example}
\squeezeexample
\squeezeexample
Consider the Consortium system (\figref{consortiumcode}). According to \algoref{init-regions}, the initial regions include $r_1 = \{(\code{Decided},\code{decision})\}$ with $\regionBound(r_1) =1$ which captures the results of the \code{vc} consensus event, and $r_2 = \{(\code{Election},\code{decision})$,
$(\code{Election},\code{motion})\}$ with $\regionBound(r_2) =1$ which captures the initial default values. 
\algoref{exp-regions} expands the initial regions. One such expansion step of region $r_1$ through edge $\edge$ from \exref{locStruc} adds the node-variable pair $(\code{Announce},\code{decision})$ to $r_1$ since the value in the \code{decision} variable is preserved along edge $\edge$.
A minimal valid \abr that meets conditions (i) through (iv) is $\region =$  
$\{(\code{Election},$ $\code{motion})$, $(\code{Election},\code{decision})$, $(\code{Decided},\code{decision})$, $(\code{Announce},$ $\code{decision})$, $(\code{LeaderDone},\code{decision})$, $(\code{ReplicaDone},\code{decision})\}$.
This $\region$ is the result of merging the expansions of $r_1$ and $r_2$ and hence has $\regionBound(\region) = max(\regionBound(r_1),\regionBound(r_2)) = 1$.
Finally, $\localBound$ is 2 since both the \code{decision} and \code{motion} variables appear in node-variable pairs outside of $\region$ (e.g., in the $(\code{Wait},\code{decision})$ and $(\code{Wait},\code{motion})$ node-variable pairs).
\end{example}

\section{Solving PMCP for Effectively Bounded $\langle \mnv,\globalSpec \rangle$}

Let $\langle \mnv, \globalSpec \rangle$ be a \reducible verification problem with domain cutoff $\boundedDomSet$.
Let $c$ be a process cutoff computable by some 
cutoff-based procedure (\eg,~\cite{Jaber.QuickSilver.2021}) 
for parameterized verification of the reduced system $\mnvb$ w.r.t. 
safety specification $\globalSpec$. 
We then refer to the tuple $\langle \mnv,\globalSpec \rangle$ as \emph{effectively bounded}. 
The procedure for parameterized verification of effectively bounded 
 $\langle \mnv,$ $\globalSpec \rangle$ is summarized by the following result. 
 \begin{theorem}
  $\forall n: \mnv \models \globalSpec \; \text{iff}  \;$ 
  $\forall n: \mnvb \models \globalSpec \;  \text{iff} \;$
  $\mathcal{M}_\Delta(c) \models \globalSpec$.
\end{theorem}

\section{Evaluation}
~\label{sec:eval}
In this section, we present the implementation of our technique and evaluate it on a set of doubly-unbounded \dab systems.

\para{\tool}. 
We build a tool, \tool, for verification of doubly-unbounded \dab systems 
that combines our domain reduction with a recent tool, \kinarach~\cite{Jaber.QuickSilver.2021}, 
for parameterized verification of \dab systems with finite-state processes. 
As detailed in \secref{value-bounding}, \tool yields a program with finite data domains, 
paving the way for \kinarach to perform its reasoning.

%
%

While each expansion step in \algoref{exp-regions} can only expand a region with a single node-variable pair, 
in \tool, we generalize an expansion step to expand a region with a strongly-connected set of node-variable pairs, even when its pairs cannot be used individually. 
Further, we emphasize that \tool automatically identifies data domains which can be treated as
scalarsets. 
%
%

\fixedskip
\para{Case Studies}.
We now demonstrate the efficacy of our technique on two case studies, Consortium and Distributed Register, that are representative of commonly used \dab systems. 

Consortium was introduced in \secref{examples} along with its safety property.
We now introduce three variants of this system. The first variant, \emph{Consortium-Three}, elects a consortium of size three instead of two. The second variant, \emph{Consortium-BCast}, does not elect a particular trusted actor to announce the decision, but rather allows either elected actor to perform a broadcast of the decided value to the rest of the system. The third variant, \emph{Consortium-Check}, forces both of the trusted actors to share the decided value, and allows the rest of the actors to check if the shared values are identical, and if not, move to an error state.

Distributed Register is based on Atomix's AtomicValue~\cite{atomix} which gives a consistent view of some stored value under concurrent updates. In Distributed Register, clients in the environment submit requests to read from and update a register. Processes service read requests from their local copy of the register.  For update requests, processes use consensus to determine a consistent value to be stored in the shared register.
The safety property is that any two processes that are in a location where they serve client read requests always have equal values in their local copy of the register. 
We define a variant of this system, called \emph{DistReg-Two}, that allows two registers to be manipulated simultaneously. 

For each case study, we also define versions, \emph{Consortium-32Bit} and \emph{DistReg-32Bit}, respectively, with 32-bit integer domains. 

\fixedskip
\para{Evaluation}.
We now discuss the result of evaluating our case studies in \tool. All experiments are performed on a MacBook Pro with Intel Core i5 CPU and 16 GB of RAM.

\begin{table}[ht]
	\squeezecaption
	\squeezecaption
	\squeezecaption
	\squeezecaption
	\centering
	\caption{Performance of \tool.}
	
	\begin{tabular}{l c c c c r}
	\hline 
	\toprule
	\multirow{2}{*}{\textbf{Benchmark\hspace{3em}}} 		& \multirow{2}{*}{\textbf{   LoC   }} 	& \textbf{ Domain } & \textbf{ Domain } & \textbf{  Process  } 	&	\multirow{2}{*}{\textbf{   Time(s)}} \\
	& & \textbf{ Size } & \textbf{ Cutoff } & \textbf{ Cutoff    } & \\
	\midrule 
Consortium			&	62		&	$\infty$	&	3	&	3	&$	0.464	\pm	0.012	 $\\
Consortium-Three	&	62		&	$\infty$	&	3	&	4	&$	1.340	\pm	0.007	 $\\
Consortium-BCast	&	58		&	$\infty$	&	3	&	3	&$	0.231	\pm	0.003	 $\\
Consortium-Check	&	68		&	$\infty$	&	3	&	5	&$	3.209	\pm	0.013	 $\\
Consortium-32Bit	&	62		&	$2^{32}$	&	3	&	3	&$	0.450	\pm	0.002	 $\\
\midrule
DistReg				&	34		&	$\infty$	&	2	&	2	&$	0.118	\pm	0.005	 $\\
DistReg-Two			&	79		&	$\infty$	&	2	&	2	&$	2.135	\pm	0.024	 $\\
DistReg-32Bit		&	34		&	$2^{32}$	&	2	&	2	&$	0.116	\pm	0.003	 $\\

	\bottomrule & 
	\end{tabular} 
	\label{table:resultss}
	\reallysqueezecaption
	
\end{table}

The performance of \tool is shown in \tabref{resultss}. For each benchmark, we provide the number of lines of code needed to model the benchmark in \aml, the initial size of the data domain (marked $\infty$ if unbounded), the domain cutoff computed by $\tool$, the process cutoff used for verification\footnote{While \tool is able to compute a domain cutoff for DistReg-Two, \kinarach is unable to compute a process cutoff. Because \tool can potentially be combined with tools beyond \kinarach for computing process cutoffs, we make an exception for this case by manually computing a process cutoff of 2.}, and the mean run time for 10 verification runs as well as the 95\% confidence intervals.

\tool successfully reduces both large and unbounded data domains 
to relatively small, bounded domains (2 and 3). 
Notably, for the DistReg-Two benchmark, \tool identifies two unbounded 
data domains (corresponding to the two registers) 
and computes domain cutoffs for each independently.
Finally, \tool is able to verify each of our benchmarks 
rather efficiently in under 4 seconds.

\section{Related Work}
\label{sec:related}

The formal methods community has developed a variety of techniques for verification of infinite-state concurrent/distributed systems. At the coarsest level, these approaches can be divided into semi-automated \cite{Gleissenthall.Pretend.Synchrony.POPL.2019,Aneris.2020,I4,Padon.PaxosMadeEPR.OOPSLA.2017,Padon.IvySafetyVerification.PLDI.2016,Taube.ModularityDecidabilityDeductive.PLDI.2018} and fully-automated \cite{Abdulla.GeneralDecidabilityTheorems.1996,abdulla2016parameterized,Abdulla.VerificationInfiniteState.2007,Alur.AutomaticCompletionDistributed.X.2015,AminofKRSV18,AusserlechnerJK16,BloemETAL15,McMillan.SymbolicModelChecking.1992,McMillan.CompositionalModelChecking.1989,Clarke.SymmetryReductions.1998,ClarkeTV06,Emerson.UtilizingSymmetry.1997,Emerson.VirtualSymmetryReduction.2000,EK00,EK03a,emerson2003combining,FS01,ip1996better,GSP,Jaber.QuickSilver.2021,Jacobs.AnalyzingGuardedProtocols.X.2018,KaiserKW10,Lubachevsky.Compact.1984,Maric.CutoffBoundsConsensus.X.2017,SchmitzS13} techniques.
Semi-automated verification typically involves manually discovering inductive invariants \cite{Padon.IvySafetyVerification.PLDI.2016,Wilox.Verdi.PLDI.2015}. Fully automated verification (the focus of this paper) requires no user intervention and is challenging due to the unbounded nature of the targeted systems.

Symbolic representations and reduction strategies are typically employed to scale or enable automated verification for systems with large or infinite state spaces. 
For instance, 
symmetry reduction \cite{Alur.AutomaticCompletionDistributed.X.2015,Clarke.SymmetryReductions.1998,Emerson.UtilizingSymmetry.1997,Emerson.VirtualSymmetryReduction.2000,emerson2003combining,ip1996better} is used to explore the state space of a system via a reduced quotient structure.
Automated verification of distributed systems is often complicated further by an unbounded number of processes, infinite data domains, or both. 
A popular approach to handling systems with an unbounded number of processes is to use (process) cutoffs (\cite{abdulla2016parameterized,AminofKRSV18,AusserlechnerJK16,BloemETAL15,ClarkeTV06,EK00,EK03a,GSP,Jaber.QuickSilver.2021,Jacobs.AnalyzingGuardedProtocols.X.2018,KaiserKW10,Maric.CutoffBoundsConsensus.X.2017}). 
Another approach for dealing with an unbounded number of processes and/or unbounded variable domains is to impose a well-order~\cite{FS01,SchmitzS13} over the state space of the system. 


In what follows, we briefly explore prior work on verification of distributed systems with unbounded processes or infinite process state spaces. 
%

\fixedskip
\para{Bounded Number of Processes, Unbounded Process States.}
One inspiration to this paper is the notion of \emph{data saturation}, first broached by \cite{Lubachevsky.Compact.1984} and explored further by \cite{ip1996better}. 
Infinite state spaces are also handled in well-ordering based frameworks \cite{Abdulla.GeneralDecidabilityTheorems.1996}
or using ``temporal case splitting'' wherein each case identifies a particular value and all other values are represented with a symbolic constant \cite{McMillan.CompositionalModelChecking.1989}.
None of these approaches target doubly-unbounded systems.

\fixedskip
\para{Unbounded Number of Processes, Bounded Process States.}
The two common approaches for parameterized verification of finite-state processes are based on 
well-ordering~\cite{EsparzaFM99,FS01,SchmitzS13} and cutoff results computed either statically \cite{AminofKRSV18,AusserlechnerJK16,ClarkeTV06,EK00,EK03a,GSP,Jacobs.AnalyzingGuardedProtocols.X.2018,Maric.CutoffBoundsConsensus.X.2017} or dynamically~\cite{abdulla2016parameterized,KaiserKW10}.
Notably, many of these cutoff results can potentially be combined with our domain reduction approach to enable verification of doubly-unbounded systems.

\fixedskip
\para{Doubly-Unbounded Systems.}
A related effort~\cite{Abdulla.VerificationInfiniteState.2007} tackles parameterized verification of infinite-state systems using a well-ordering over global states and performing a backward-reachability analysis. Our approach differs from this effort in the target distributed systems and their models, as well as the technical approach.
The effort in \cite{Abdulla.VerificationInfiniteState.2007} targets generic distributed systems that are modeled using the more-traditional system model where processes are defined as extended finite-state automata with local variables and global conditions on transitions. On the other hand, we target modularly-designed \dab systems built with abstractions of agreement protocols and, in particular, modeled in an {\em easy-to-use} modeling language \aml. 
Further, the reduction in~\cite{Abdulla.VerificationInfiniteState.2007} does not require systems to be value-symmetric and tackles the unboundedness of \emph{both} the number of processes and the process state space in one {\em consolidated} step. In contrast, while the reduction in our approach requires the system to be value-symmetric, it is {\em separable}, and, arguably, more {\em flexible}. 
Specifically, our approach relies on two separate reductions, one for the process state spaces and one for the number of processes, each of which can potentially be replaced with other reductions. 

The authors of \cite{ClarkeTV06} combine predicate and counter abstractions into an ``environment abstraction'' to verify doubly-unbounded systems. The environment abstraction designates one process as the ``reference'' process and models all other processes in relation to it. 
Our work differs from their approach in two ways: (i) they do not verify systems which incorporate abstractions of agreement protocols, and (ii) they capture the relationships between values in the system using predicates describing the environment of the reference process, while we capture such relationships during our bounded region analysis.

The authors of \cite{Maric.CutoffBoundsConsensus.X.2017} reduce parameterized verification of consensus algorithms over inputs from infinite domains to parameterized verification of these algorithms over binary inputs. They provide a process cutoff to reduce the problem to finite verification of consensus algorithms over binary inputs. Aside from focusing on consensus algorithms (as opposed to  systems built upon them), their work also differs from ours in that their reduction of infinite domains to finite ones relies on the zero-one principle for sorting networks rather than a static analysis of the system being verified, like our bounded region analysis.

\newpage 

\bibliographystyle{splncs04}
\bibliography{main}

\newpage 

\appendix
\section{Formal Semantics}
\label{app:semanticsApp}

In this section, we present formal definitions of the local and global semantics of \aml program definitions, as presented in figures \figref{localSemRules} and \figref{globalSemRules}, respectively.

\begin{figure}[ht]
\centering
\begin{tcolorbox}[colback=white,sharp corners,boxrule=0.3mm,top=1mm,bottom=1mm, left=1mm]
\begin{minipage}{0.49\textwidth}
\scriptsize
$\inferrule[broadcast send]
{
    \action = \acting{\evSigTuple{\actionName}{\val}{\envFlag}{\bcCoordType}} \in [\acting{\eventSig}] \\
    \localState(var_{\acting{\eventSig}}) = \val \\
    \exists (\localGuard, \acting{\eventSig}, \updates)\in \handlerSet. \localState\models \phi \land \updates(\localState) =  \localState'
}
{(\localState,\action, \localState') \in  \localTransRel}$\\ \\

$\inferrule[rendezvous send]
{
    \action = \acting{\evSigTuple{\actionName}{\val}{\envFlag}{\pwCoordType}} \in [\acting{\eventSig}] \\
    \localState(var_{\acting{\eventSig}}) = \val \\
    \exists (\localGuard, \acting{\eventSig}, \updates) \in \handlerSet. \localState \models \phi \land \updates(\localState) = \localState'
}
{(\localState, \action, \localState') \in \localTransRel}$\\ \\

$\inferrule[partition win]
{
    \action = \acting{\evSigTuple{\actionName}{\winset}{\envFlag}{\pcCoordType}} \in [\acting{\eventSig}] \\
    \localState(self) \in \winset \\
    \exists (\localGuard, \acting{\eventSig}, \updates) \in \handlerSet. \localState_i\models \phi \land \updates(\localState) = \localState'
}
{(\localState, \action, \localState') \in  \localTransRel}$\\ \\

$\inferrule[partition lose]
{
    \action = \reacting{\evSigTuple{\actionName}{\winset}{\envFlag}{\pcCoordType}} \in [\reacting{\eventSig}] \\
    \localState(self) \notin \winset \\
    \exists (\localGuard, \reacting{\eventSig}, \updates) \in \handlerSet. \localState_i\models \phi \land \updates(\localState) = \localState'
}
{(\localState, \action, \localState') \in  \localTransRel}$\\ \\

$\inferrule[internal]
{
    \exists (\localGuard, \epsilon, \updates) \in \handlerSet. \localState\models \phi \land \updates(\localState) = \localState'
}
{(\localState, \epsilon, \localState') \in  \localTransRel}$ \\  \\

\end{minipage}
\begin{minipage}{0.49\textwidth}
\scriptsize
$\inferrule[broadcast receive]
{
    \action = \reacting{\evSigTuple{\actionName}{\val}{\envFlag}{\bcCoordType}} \in [\reacting{\eventSig}] \\
    \exists (\localGuard, \reacting{\eventSig}, \updates) \in \handlerSet. \localState\models \phi \land \updates[\val/var_{\reacting{\eventSig}}](\localState) = \localState'
}
{(\localState, \action, \localState') \in  \localTransRel}$\\ \\

$\inferrule[rendezvous receive]
{
    \action = \reacting{\evSigTuple{\actionName}{\val}{\envFlag}{\pwCoordType}} \in [\reacting{\eventSig}] \\
    \exists (\localGuard, \reacting{\eventSig}, \updates) \in \handlerSet. \localState\models \phi \land \updates[\val/var_{\reacting{\eventSig}}](\localState) = \localState'
}
{(\localState, \action, \localState') \in  \localTransRel}$\\ \\

$\inferrule[Value Consensus (acting)]
{
    \action = \acting{\evSigTuple{\actionName}{\val}{\envFlag}{\vcCoordType}} \in [\acting{\eventSig}] \\
    \localState(var_{\acting{\eventSig}}) \in \val \\
    \exists (\localGuard, \acting{\eventSig}, \updates) \in \handlerSet. \localState\models \phi \land \updates[\val/var_{\reacting{\eventSig}}](\localState) = \localState'
}
{(\localState, \action, \localState') \in  \localTransRel}$\\ \\

$\inferrule[Value Consensus (reacting)]
{
    \action = \reacting{\evSigTuple{\actionName}{\val}{\envFlag}{\vcCoordType}} \in [\reacting{\eventSig}] \\
    \exists (\localGuard, \reacting{\eventSig}, \updates) \in \handlerSet. \localState\models \phi \land \updates[\val/var_{\reacting{\eventSig}}](\localState) = \localState'
}
{(\localState, \action, \localState') \in  \localTransRel}$\\ \\

$\inferrule[Update]
{
    \forall (\lhs, \rhs) \in \updates. \localState'(\lhs) = \rhs\\
    \forall \varDef \in \varDefSet. \neg\exists (\varName, \rhs) \in \updates \Rightarrow \localState(\varName) = \localState'(\varName)
}
{\updates(\localState) = \localState'}$
\end{minipage}
\end{tcolorbox}

\caption{Local operational semantics of \aml programs.}
\label{fig:localSemRules} 
\end{figure}

In addition to the broadcast and consensus transitions defined in \secref{mercury}, we formally present the remaining transition types below.

\begin{compactenum}[1.]
	\item $\globalTransRel$ contains an internal global transition can occur when one process has a local internal transition and all other processes remain in the same local states,
	
	\item a \pairwise transition for event $\eventName$ with coordination type $\pwCoordType$ can occur when one process has a local transition for $\acting{\eventName}$, one other process has corresponding $\reacting{\eventName}$ local transition, and all other processes remain in the same local states,
	
	\item $\globalTransRel$~contains a ``\PartitionCons transition'' $(\globalState, \event, \globalState')$ for partition event $\event$
		with participant set $\participantSet$
	and winning set $\winSet$ iff 
		(1) each participating process in $\participantSet$ has a consistent view of the other participants by holding $\participantSet$ in a local participant set variable,
	(2) each process $\process_i$ in $\winSet$ has a local ``partition win'' transition $(\globalState[i], \acting{\event},\globalState'[i])$, 
	(3) each process  $\process_j$ in $\participantSet \setminus \winSet$ has a corresponding local ``partition lose'' transition $(\globalState[j], \reacting{\event},\globalState'[j])$, and 
	(4) the local states of all other processes remain unchanged.
	 %
	 %
	  
\end{compactenum}

\begin{figure}[ht]
\centering
\begin{tcolorbox}[colback=white,sharp corners,boxrule=0.3mm,top=1mm,bottom=1mm, left=1mm]
\begin{minipage}{\textwidth}
\centering
\scriptsize
$\inferrule[internal]
{
    \exists i: (\globalState[i], \intCoordType, \altGlobalState[i]) \in  \localTransRel\\
    \forall j \neq i: \altGlobalState[j] = \globalState[j]
}
{(\globalState, \intCoordType, \altGlobalState) \in \globalTransRel}$
 \\\phantom{lol}\\ 
$\inferrule[broadcast]
{
    \event = (\actionName, \val, \bcCoordType, \envFlag)\\
    \exists i:(\globalState[i], \acting{\event}, \altGlobalState[i]) \in \localTransRel\\
    \forall j \neq i: (\globalState[j], \reacting{\event}, \altGlobalState[j]) \in \localTransRel
}
{(\globalState, \event, \altGlobalState) \in \globalTransRel}$
 \\\phantom{lol}\\
$\inferrule[rendezvous]
{
    \event = (\actionName, \val, \pwCoordType, \envFlag)\\
    \exists i:(\globalState[i], \acting{\event}, \altGlobalState[i]) \in \localTransRel\\
    \exists j:(\globalState[j], \reacting{\event}, \altGlobalState[j]) \in \localTransRel\\
    \forall k: k \notin \{i, j\} \implies \altGlobalState[k] = \globalState[k]
}
{(\globalState, \event, \altGlobalState) \in \globalTransRel}$
 \\\phantom{lol}\\
$\inferrule[partition]
{
    \event = (\actionName, \winset, \pcCoordType, \envFlag)\\
    \exists S \subseteq I_n: (\winset \subseteq S \land 
    \forall i \in S: q[i].\sigma(\text{ptct}_\event) = S) \land  
    (|\winset| = \text{card}_\event \lor S = \winset)\\
    \forall i \in \winset: (\globalState[i], \acting{\event}, \altGlobalState[i]) \in \localTransRel\\
    \forall j \in S \setminus \winset: (\globalState[j], \reacting{\event}, \altGlobalState[j]) \in \localTransRel\\
    \forall k \in I_n \setminus S: \altGlobalState[k] = \globalState[k]
}
{(\globalState, \event, \altGlobalState) \in \globalTransRel}$
 \\\phantom{lol}\\
$\inferrule[consensus]
{
    \event = (\actionName, \val, \vcCoordType, \envFlag) \in [\eventSig]\\
    \exists S \subseteq I_n: \forall i \in S: \globalState[i](\text{ptct}_\eventSig) = S\\
    \forall w \in \val: \exists i \in S: \globalState[i](var_{\acting{\eventSig}}) = w\\
    \forall j \in S: (\globalState[j], \reacting{\event}, \altGlobalState[j]) \in \localTransRel\\
    \forall j \in I_n \setminus S: \altGlobalState[j] = \globalState[j]
}
{(\globalState, \event, \altGlobalState) \in \globalTransRel}$
\end{minipage}
\end{tcolorbox}

\caption{Global operational semantics of \aml programs.}
\label{fig:globalSemRules} 
\end{figure}
\section{Proofs}
\label{app:proofs}

\subsection{Local Transitions under Permutations}
\label{app:proofs-trans-local}

\begin{lemma}
\label{lem:local-trans-preserved-app}
$\forall \localPerm: \dom \mapsto \dom, (\localState,\ \action,\ \localState') \in  \localTransRel:
\localPerm((\localState,\ \action,\ \localState')) \in \localTransRel
$
\end{lemma}

\begin{proof}
We examine an arbitrary permutation $\localPerm$ over $\dom$ and an arbitrary local transition $(\localState,\ \action,\ \localState') \in  \localTransRel$. By the definition of scalarsets, values in $\dom$ may only be compared with (dis)equality, so the satisfaction of any local guard $\localGuard$ associated with a handler is invariant to permutation; i.e. $\localState \models \localGuard \implies \localPerm(\localState) \models \localGuard$. Additionally, since all updates $\updates$ involving values from $\dom$ must be direct assignments, it follows immediately that $\updates(\localState) = \localState' \implies \updates(\localPerm(\localState)) = \localPerm(\localState')$.
Finally, since $\localPerm$ maps values in $\dom$ to values in $\dom$, it is trivial that for any action $\action = \direction{(\actionName,\ \val,\ \coordType ,\ \envFlag)}$ in $[\direction{\eventSig}]$, the permuted action $\localPerm(\action) = \direction{(\actionName,\ \localPerm(\val),\ \coordType ,\ \envFlag)}$ is also in $[\direction{\eventSig}]$. So, by the appropriate local operational semantics rule in \figref{localSemRules}, $\localPerm((\localState,\ \action,\ \localState')) = (\localPerm(\localState),\ \localPerm(\action),\ \localPerm(\localState')) \in \localTransRel$.
\end{proof}

\subsection{Domain Reduction Congruence}
\label{app:backward-congruence}

\begin{lemma}
\label{lem:backward-congruence-app}
$
\forall \altGlobalState \in \globalStateSet, \boundedAltGlobalState \in \bounded{\globalStateSet}, 
(\globalState, \event, \altGlobalState) \in \globalTransRel:
\altGlobalState \cwpRel \boundedAltGlobalState
\implies
(\exists (\boundedGlobalState, \boundedEvent, \boundedAltGlobalState) \in \bounded{\globalTransRel}:
\globalState \cwpRel \boundedGlobalState).
$

\end{lemma}
Since $ \altGlobalState \cwpRel \boundedAltGlobalState$, we know there must exist a CWP $\globalPerm'$ such that $\globalPerm'(\altGlobalState) = \boundedAltGlobalState$.
To prove the lemma, we must identify an appropriate global state $\boundedGlobalState$ and $\boundedEvent$ such that $(\boundedGlobalState, \boundedEvent, \boundedAltGlobalState) \in \bounded{\globalTransRel}$ and $\globalState \cwpRel \boundedGlobalState$. Since $\boundedGlobalState$ must be related to $\globalState$ by $\cwpRel$, we need to show that there exists another CWP $\globalPerm \in \cwpSet_\dom^\boundedRegion(\globalState)$ such that $\minimal(\globalPerm,\globalState)$ and  $\globalPerm(\globalState) = \boundedGlobalState$. By identifying a $\globalPerm$ and $\boundedEvent$ such that $(\globalPerm(\globalState), \boundedEvent, \boundedAltGlobalState) \in \bounded{\globalTransRel}$, the lemma is proven by letting $\boundedGlobalState = \globalPerm(\globalState)$.
We carefully define $\globalPerm$ to (i) agree with $\globalPerm'$ on values preserved in the transition from $\globalState$ to $\altGlobalState$ so that $\globalPerm(\globalState)$ may transition to $\boundedAltGlobalState$, (ii) permute values in $\projection_\dom^\boundedRegion(\globalState)$ consistently and values outside of $\projection_\dom^\boundedRegion(\globalState)$ to the smallest available values
 so that $\globalPerm \in \cwpSet_\dom^\boundedRegion(\globalState)$.
More precisely, we detail a procedure $\mkperm$ below which, given $\globalPerm'$, $\globalState$, and $\altGlobalState$, constructs the CWP $\globalPerm = \mkperm(\globalPerm', \globalState, \altGlobalState)$ as follows:

\begin{compactenum}
\item For any two process indices $i, j \in [1, n]$ and value $\val \in \dom$, if $\val$ is in both $\projection_\dom^\boundedRegion(\globalState[i])$
and $\projection_\dom^\boundedRegion(\altGlobalState[j])$,
then for each such $i$, let $\globalPerm[i]$ permute $\val$ to the same value as the permuted value of $\val$ in $\globalPerm'[j]$. That is,
 \begin{multline*}
\forall i, j \in [1, n], \val \in \dom: 
\val \in \projection_\dom^\boundedRegion(\globalState[i])
\land  \val \in \projection_\dom^\boundedRegion(\altGlobalState[j])
\implies \\ \globalPerm[i](\val) = \globalPerm'[j](\val).
\end{multline*}

\item For each process indices $i,j \in [1, n]$ and variables $\varName,\varName'$, if $\getsFrom(\altGlobalState[j], \varName', \globalState[i], \varName)$ then let $\globalPerm[i]$ permute $\globalState[i](\varName)$ to the same value as the permuted value of $\altGlobalState[j](\varName')$ in $\globalPerm'[j]$. That is, 
\begin{multline*}
\forall i,j \in [1, n], \varName,\varName'\in \vars: \\
\getsFrom(\altGlobalState[j], \varName', \globalState[i], \varName) \land \altGlobalState[j](\varName') = \globalState[i](\varName) \implies \\ \globalPerm[i](\globalState[i](\varName)) = \globalPerm'[j](\altGlobalState[j](\varName')).
\end{multline*}

\item For any value $\val \in \dom$ for which there exists some process index $i \in [1,n]$ such that
$\val \in \projection_\dom^\boundedRegion(\globalState[i])$
but there does not exist a process index $j \in [1, n]$ such that $\val \in \projection_\dom(\altGlobalState[j])$,
then for all $i \in [1,n]$ let $\globalPerm[i]$ for each process index $i$ permute $\val$ to the smallest value $\val'$ (according to $\domainOrder$) that was not used in any of the previous steps (for any process).

\item For each process $i \in [1, n]$ and variable $\varName \in \vars$ of type $\dom$,
if there does not exist another process $j \in [1, n]$ and variable $\varName'$ such that $\getsFrom(\altGlobalState[j], \varName', \globalState[i], \varName)$ let $\globalPerm[i]$ permute $\globalState[i](\varName)$ to the smallest value $\val'$ (according to $\domainOrder$) such that no value from steps 1 or 3 was permuted to $\val'$ in $\globalState[i]$.

\item For each $i \in [1, n]$, permute any remaining values to the next value $\val'$ (according to $\domainOrder$) that was not used in any of the previous steps (for any process).
%
\end{compactenum}

\ourskip
\para{Properties of $\mathbf{mk\boldsymbol{\globalPerm}(\boldsymbol{\globalPerm}', \globalState, \altGlobalState)}$}.
The above construction
gives the CWP $\globalPerm = \mkperm(\globalPerm', \globalState, \altGlobalState)$ several interesting properties. The first of which is that a value which stays in the system during the transition (by being retained in a variable) is permuted consistently in $\globalPerm$ and $\globalPerm'$:
\begin{multline*}
	\forall i \in [1, n], \varName \in \vars: \exists j \in [1, n], \varName' \in \vars: \\
	\getsFrom(\altGlobalState[j], \varName', \globalState[i], \varName) \implies \mkperm(\globalPerm', \globalState, \altGlobalState) [i](\globalState[i](\varName)) = \globalPerm'[j](\altGlobalState[j](\varName'))
	\tag{$\globalPerm$-kept}
\end{multline*}
Another property of the CWP $\globalPerm$ is that any value in $\projection_\dom^\boundedRegion(\globalState)$ is permuted consistently by all components of $\globalPerm$:
\begin{multline*}
	\forall i, j \in [1, n], \val \in \projection_\dom^\boundedRegion(\globalState): \\ \mkperm(\globalPerm', \globalState, \altGlobalState) [i](\val) = \mkperm(\globalPerm', \globalState, \altGlobalState) [j](\val)
	\tag{$\globalPerm$-stable}
\end{multline*}
The construction also ensures that $\globalPerm$ is a bijection:
\begin{multline*}
	\forall i \in [1, n], \val, \val' \in \dom: \val = \val' \iff \\
	\mkperm(\globalPerm', \globalState, \altGlobalState) [i](\val) = \mkperm(\globalPerm', \globalState, \altGlobalState) [i](\val')
	\tag{$\globalPerm$-bijection}
\end{multline*}

Ultimately, we are able to conclude that the permutation $\globalPerm$ maps the values of $\globalState$ into the bounded range of $\bounded{\dom}$.
\begin{multline*}
	\globalPerm'(\altGlobalState) \in \bounded{\globalStateSet} \implies 
	\forall i \in [1, n], \val \in \dom: \mkperm(\globalPerm', \globalState, \altGlobalState) [i](\val) \in \domainOrder[1, y]
	\tag{$\globalPerm$-bounded}
\end{multline*}
where $y = \regionBound + \localBound$ is the size of the reduced scalarset domain $\bounded{\dom}$, and $\domainOrder[1, y]$ is the set of the $y$ most minimal values according to $\domainOrder$.

This property comes from the assumption that there are no more than $\regionBound$ values $\val$ in $\projection_\dom^\boundedRegion(\altGlobalState[i])$
for any process $i$,
and any of them which are also in $\projection_\dom^\boundedRegion(\globalState)$ will be permuted to the same values as $\globalPerm'(\altGlobalState)$ (all of which are in $\domainOrder[1, y]$, since $\globalPerm'(\altGlobalState) = \boundedAltGlobalState \in \bounded{\globalStateSet}$).

Then since each $\altGlobalState[j]$ may only hold up to $\localBound$ values which are not stable according to any process (i.e. $\val \notin \projection^\boundedRegion_\dom(\globalState)$),
these values are all permuted in $\boundedAltGlobalState$ to a particular set of $\localBound$ values in $\domainOrder[1, y]$ (by our minimality constraint), and our assumptions prevent more than 1 process $j$ from getting a value $\val \notin \projection^\boundedRegion_\dom(\globalState)$ from $\globalState[i]$,
so for some variable(s) $\varName$ and $\varName'$, if $\globalState[i](\varName) = \val$ and $\getsFrom(\altGlobalState[j], \varName', \globalState[i], \varName)$, $\globalPerm[i]$ permutes $\val$ to the same single value as $\globalPerm'[j]$ permutes it, which is in the set of $\localBound$ values in $\boundedAltGlobalState$ to which unstable values in $\boundedGlobalState$ are permuted.

Then, if there are only $\regionBound' < \regionBound$ values $\val$ in $\globalState[i]$ such that $\val \in \projection^\boundedRegion_\dom(\globalState)$,
there must be $\regionBound - \regionBound'$ values in $\domainOrder[1, y]$ such that no value has been permuted to them in the first two steps, any additional values $\val \in \projection^\boundedRegion_\dom(\globalState)$, of which there may be at most $\regionBound - \regionBound'$, are permuted to the next available value according to $\domainOrder$ (which must necessarily be in $\domainOrder[1, y]$, otherwise there could not have been only $\regionBound'$ values $\val \in \projection^\boundedRegion_\dom(\globalState)$.

Lastly, if there are only $\localBound' < \localBound$ values $\val \in \projection_\dom(\globalState[i]) \setminus 
\projection_\dom^\boundedRegion(\globalState)$ for $i \in [1, n]$, 
there must be at most $\localBound - \localBound'$ values in 
$\domainOrder[1, y]$ such that no value has been permuted to them in $\globalPerm[i]$ by the first three steps, any additional values $\val \in \projection_\dom(\globalState[i]) \setminus 
\projection_\dom^\boundedRegion(\globalState)$,
of which there may be at most $\localBound - \localBound'$, are permuted to the next available value according to $\domainOrder$
which no value in $\globalState[i]$ has been permuted yet (and which must necessarily be in $\domainOrder[1, y]$, otherwise there could not have been only $\localBound'$ values in $ \projection_\dom(\globalState[i]) \setminus 
\projection_\dom^\boundedRegion(\globalState)$.

Thus, $\forall i \in [1, n], \val \in \dom: \mkperm(\globalPerm', \globalState, \altGlobalState) [i](\val) \in \domainOrder[1, y]$, which leads us to the following conclusion:
\[
\globalPerm'(\altGlobalState) \in \bounded{\globalStateSet} \implies \mkperm(\globalPerm', \globalState, \altGlobalState)(\globalState) \in \bounded{\globalStateSet}
\tag{$\globalPerm$-bounded-upshot}
\]

\para{Transition Case Analysis}
With such $\globalPerm$ in hand, it remains to show that there exists an event $\boundedEvent$ such that the transition $(\globalPerm(\globalState), \boundedEvent, \boundedAltGlobalState) $ is a valid transition in $ \bounded{\globalTransRel}$. In what follows, we perform a case analysis over all types of events $\event$ and show, for each type of global transition $(\globalState, \event, \altGlobalState) \in \globalTransRel$, that such an event $\boundedEvent$ exists.

\begin{lemma}
	$\exists \globalPerm, \boundedEvent:
	(\globalPerm(\globalState), \boundedEvent, \boundedAltGlobalState) \in \bounded{\globalTransRel}$
\end{lemma}
\begin{proof}
	\phantom{nextline}

	In the case that $\event$ is an internal coordination event, we prove the lemma by contradiction, as follows.
	
	\begin{footnotesize}
		\begin{tabular}{r r l c r}
	
	1. & & $(\globalState, \event, \altGlobalState) \in \globalStateSet	$ &   & (assumption)  \\ 
	2. & & $\globalPerm'(\altGlobalState) = \boundedAltGlobalState \in \bounded{\globalStateSet}$ &   & (assumption) \\ 
	3. & & $\altGlobalState \cwpRel \boundedAltGlobalState$ &   & (assumption) \\ 
	4. & & $\neg\exists \globalPerm, \boundedEvent: (\globalPerm(\globalState), \boundedEvent, \boundedAltGlobalState) \in \bounded{\globalTransRel}$ &   & (assumption) \\ 
	5. & & $e = \text{in}$ &   & (assumption) \\ 
	6. & & $\exists i': (\globalState[i'],\ \text{in},\ \altGlobalState[i'])\ \in \ \localTransRel \land$ &   &  \\ 
	& & $\forall j' \neq i': \altGlobalState[j']\ =\ \globalState[j']$ &   & (1, 5) \\ 
	7. & & $(\globalState[i],\ \text{in},\ \altGlobalState[i])\ \in \ \localTransRel \land \forall j' \neq i: \altGlobalState[j']\ =\ \globalState[j']$ &   & (6, $\exists i' = i$ (arbitrary)) \\ 
	8. & & $(\globalState[i],\ \text{in},\ \altGlobalState[i])\ \in \ T$ &   & (7, $\land$) \\ 
	9. & & $\forall j' \neq i: \altGlobalState[j']\ =\ \globalState[j']$ &   & (7, $\land$) \\ 
	10. & & $\forall \globalPerm, \boundedEvent:(\globalPerm(\globalState), \boundedEvent, \boundedAltGlobalState) \notin \bounded{\globalTransRel}$ &   & (4, $\neg \exists$) \\ 
	11. & & $\forall \globalPerm: (\globalPerm(\globalState), \text{in}, \boundedAltGlobalState) \notin \bounded{\globalTransRel}$ &   & (10, $\boundedEvent = \text{in}$) \\ 
	12. & & $\globalPerm = mk\globalPerm(\globalPerm', q, \altGlobalState) \ \land\ (\globalPerm(\globalState), \text{in}, \boundedAltGlobalState) \notin \bounded{\globalTransRel}$ &   & (11, $\forall \globalPerm = mk\globalPerm(\globalPerm', \globalState, \altGlobalState)$) \\ 
	13. & & $\globalPerm = mk\globalPerm(\globalPerm', \globalState, \altGlobalState)$ &   & (12, $\land$) \\ 
	14. & & $(\globalPerm(\globalState), \text{in}, \boundedAltGlobalState) \notin \bounded{\globalTransRel}$ &   & (12, $\land$) \\ 
	15. & & $\forall \varName' \in \vars: \exists \varName \in \vars:$ &   & \\ 
	& & $\getsFrom(\altGlobalState[i], \varName', \globalState[i], \varName)$ &   & (8, $kept$) \\ 
	16. & & $\neg \exists i': (\globalPerm(\globalState)[i'],\ \text{in},\ \boundedAltGlobalState[i']) \in \localTransRel \land \forall j' \neq i':$ &   &  \\ 
	& & $ \boundedAltGlobalState[j'] =\globalPerm(\globalState)[j']\land\globalPerm(\globalState) \in \bounded{\globalStateSet} \land \boundedAltGlobalState \in \bounded{\globalStateSet}$ &   & (14, internal-global) \\ 
	17. & & $\forall i': (\globalPerm(\globalState)[i'],\ \text{in},\ \boundedAltGlobalState[i']) \in \localTransRel \lor \exists j' \neq i': $ &   &  \\ 
	& & $ \boundedAltGlobalState[j'] \neq \globalPerm(\globalState)[j']\lor\globalPerm(\globalState) \notin \bounded{\globalStateSet} \lor \boundedAltGlobalState \notin \bounded{\globalStateSet}$ &   & (16, DeMorgan's) \\ 
	18. &  a) & $\boundedAltGlobalState \notin \bounded{\globalStateSet}$ &   & (17, $\lor$) \\ 
	19. &  a) & $\bot$ &   & (2, 18a) \\ 
	20. &  b) & $\globalPerm(\globalState) \notin \bounded{\globalStateSet}$ &   & (17, $\lor$) \\ 
	21. &  b) & $\globalPerm(\globalState) \in \bounded{\globalStateSet}$ &   & (2, bounded-upshot) \\ 
	22. &  b) & $\bot$ &   & (20b, 21) \\ 
	23. & & $\forall i': (\globalPerm(\globalState)[i'],\ \text{in},\ \boundedAltGlobalState[i']) \in \localTransRel\ \lor $ &   & \\ 
	& & $\exists j' \neq i': \boundedAltGlobalState[j'] \neq \globalPerm(\globalState)[j']$ &   & (17, 18-19a, 20-22b) \\ 
	24. & & $(\globalPerm(\globalState)[i],\ \text{in},\ \boundedAltGlobalState[i]) \in \localTransRel \lor \exists j' \neq i: \boundedAltGlobalState[j'] \neq \globalPerm(\globalState)[j']$ &   & (23, $\forall i' = i$ (from 7)) \\ 
	25. &  a) & $\exists j' \neq i: \boundedAltGlobalState[j'] \neq \globalPerm(\globalState)[j']$ &   & (24, $\lor$) \\ 
	26. &  a) & $j \neq i \land \boundedAltGlobalState[j] \neq \globalPerm(\globalState)[j]$ &   & (25, $\exists j' = j$ (arbitrary)) \\ 
	27. &  a) & $j \neq i$ &   & (26, $\land$) \\ 
	28. &  a) & $\boundedAltGlobalState[j] \neq \globalPerm(\globalState)[j]$ &   & (26, $\land$) \\ 
	29. &  a) & $\globalPerm'(\altGlobalState)[j] \neq \globalPerm(\globalState)[j]$ &   & (28, 2) \\ 
	30. &  a) & $\forall \varName \in \vars: j' \neq i: \getsFrom(\altGlobalState[j'], \varName, \globalState[j'], \varName)$ &   & (9, internal-global) \\ 
	31. &  a) & $\forall \varName' \in \vars: \getsFrom(\altGlobalState[j], \varName', \globalState[j], \varName)$ &   & (30, $\forall j' = j$ (from 26)) \\ 
	32. &  a) & $\altGlobalState[j]\ =\ \globalState[j]$ &   & (9, 27) \\ 
	33. &  a) & $\forall \varName' \in \vars: \globalPerm[j](\globalState[j](\varName)) = \globalPerm'[j](\altGlobalState[j](\varName))$ &   & (31, $\globalPerm$-kept) \\ 
	34. &  a) & $\forall \varName' \in \vars: \globalPerm(\globalState)[j](\varName) = \globalPerm'(\altGlobalState)[j](\varName)$ &   & (33, $\globalPerm(\globalState)[i] = \globalPerm[i](\globalState[i])$) \\ 
	35. &  a) & $\globalPerm(\globalState)[j] = \globalPerm'(\altGlobalState)[j]$ &   & (34, all vars equal) \\ 
	36. &  a) & $\globalPerm'(\altGlobalState)[j] \neq \globalPerm(\globalState)[j]$ &   & (28, 2) \\ 
	37. &  a) & $\bot$ &   & (35a, 36a) \\ 
	38. &  b) & $(\globalPerm(\globalState)[i],\ \text{in},\ \boundedAltGlobalState[i]) \in \localTransRel$ &   & (24, $\lor$) \\ 
	39. &  b) & $(\globalPerm(\globalState)[i],\ \text{in},\ \globalPerm'(\altGlobalState)[i]) \in \localTransRel$ &   & (38, 2) \\
	
\end{tabular}

\begin{tabular}{r r l c r}

40. &  b) & $(\globalPerm[i](\globalState[i]),\ \text{in},\ \globalPerm'[i](\altGlobalState[i])) \in \localTransRel$ &   & (39, $\globalPerm(\globalState)[i] = \globalPerm[i](\globalState[i])$) \\ 
	41. &  b) & $(\globalPerm[i](\globalState[i]),\ \text{in},\ \globalPerm[i](\altGlobalState[i]))\ \in \ T$ &   & (8, $\pi$ preserves $T$) \\ 
	42. &  b) & $\forall \varName \in \vars: \exists j' \in [1, n], \varName' \in \vars:$ &   &  \\ 
	&   & $\getsFrom(\altGlobalState[j], \varName', \globalState[i], \varName)$ &   & (8, internal-local) \\ 
	43. &  b) & $\forall \varName' \in \vars: \globalPerm[i](\globalState[i](\varName)) = \globalPerm'[i](\altGlobalState[i](\varName))$ &   & (42, $\globalPerm$-kept) \\ 
	44. &  b) & $\globalPerm(\globalState)[i] = \globalPerm'(\altGlobalState)[i]$ &   & (43, all vars equal) \\ 
	45. &  b) & $\globalPerm[i](\globalState[i]) = \globalPerm'[i](\altGlobalState[i])$ &   & (44, $\globalPerm(\globalState)[i] = \globalPerm[i](\globalState[i])$) \\ 
	46. &  b) & $\forall \varName \in \vars: \exists \varName' \in \vars:\getsFrom(\altGlobalState[i], \varName', \globalState[i], \varName)$ &   & (8, internal-local) \\ 
	47. &  b) & $\forall \varName \in \vars: \globalPerm[i](\altGlobalState[i](\varName)) = \globalPerm'[i](\altGlobalState[i](\varName))$ &   & (46, $\globalPerm$-kept, $\getsFrom$) \\ 
	48. &  b) & $\globalPerm[i](\altGlobalState[i]) = \globalPerm'[i](\altGlobalState[i])$ &   & (47, all vals equal) \\ 
	49. &  b) & $(\globalPerm[i](\globalState[i]),\ \text{in},\ \globalPerm'[i](\altGlobalState[i]))\ \in \ T$ &   & (41, 48) \\ 
	50. &  b) & $\bot$ &   & (49, 40)
	
\end{tabular} 

All branches closed.   
	\end{footnotesize}
	
	The proof is almost identical when $\event$ is a partition event, or any interaction with the environment, as in these transitions do not transmit values between system processes. As such, we elide the proof details for these transitions here.
	
	For the remaining cases, we make two observations which will aid in our proof effort. The first is that when two local permutations treat the right-hand-side values of all assignments (i.e., those being assigned from) in a set of updates identically, the local state resulting from that set of updates is identical, regardless of which permutation is applied to the assigned values.
	\begin{multline*}
		\forall \pi, \pi', U, s, s':
		U(s) = s'
		\land (\forall v, v' \in U:
		\pi(s(v)) = \pi'(s'(v))) \\
		\implies U(\pi(s)) = \pi'(s')
		\tag{update-equiv}
	\end{multline*}

 The second is that, during a local transition, when a value moves from one variable $v$ in some state $s$ to another variable $v'$ in another state $s'$, if two permutations $\pi$ and $\pi'$ agree on the permutation of the value of $s(v)$, and the transition is preserved when both $s$ and $s'$ are permuted by $\pi$, then the transition is preserved when the source state $s$ is permuted by $\pi'$ instead of $\pi$.
\begin{multline*}
	\forall \pi, \pi':
	(\forall s, v, s', v': getsFrom(s', v', s, v) \implies \pi(s(v)) = \pi'(s(v)))
	\implies \\
	((\pi(s), e, \pi(s')) \in T
	\iff
	(\pi'(s), e, \pi(s')) \in T)
	\tag{src-equiv}
\end{multline*}

	Next, we consider the case in which $\event$ is a pairwise transmission between two non-environment processes. Again, we prove the lemma by contradiction, as follows.
	
	\begin{footnotesize}
		\begin{tabular}{r r l c r}
	
	1. &  & $(\globalState, \event, \altGlobalState) \in \globalTransRel$ &   & (assumption) \\
	2. &  & $\globalPerm'(\altGlobalState) = \boundedAltGlobalState \in \bounded{\globalTransRel}$ &   & (assumption) \\
	3. &  & $\altGlobalState \cwpRel \boundedAltGlobalState$ &   & (assumption) \\
	4. &  & $\neg\exists \globalPerm, \boundedEvent: (\globalPerm(\globalState), \boundedEvent, \boundedAltGlobalState) \in \bounded{\globalTransRel}$ &   & (assumption) \\
	5. &  & $\event = (a,\ \val,\ \text{pw},\ \bot)$ &   & (assumption) \\
	6. &  & $\exists i': (\globalState[i'],\ \event!,\ \altGlobalState[i'])\ \in \ \localTransRel \land$ &   &  \\
	&  & $\exists j': (\globalState[j'],\ \event?,\ \altGlobalState[j'])\ \in \ \localTransRel \land $ &   &  \\
	&  & $\forall k' \notin \{i', j'\} \neq i': \altGlobalState[k']\ =\ \globalState[k']$ &   & (1, pw-global) \\
	
\end{tabular}

\begin{tabular}{r r l c r}

7. &  & $(\globalState[i],\ \event!,\ \globalState'[i])\ \in \ \localTransRel \land$ &   &  \\
	&  & $(\globalState[j],\ \event?,\ \globalState'[j])\ \in \ \localTransRel \land$ &   &  \\
	&  & $\forall k' \notin \{i, j\} \neq i': \altGlobalState[k']\ =\ \globalState[k']$ &   & (6, $\exists i' = i, j' = j$) \\
	8. &  & $(\globalState[i],\ \event!,\ \globalState'[i])\ \in \ \localTransRel$ &   & (7, $\land$) \\
	9. &  & $(\globalState[j],\ \event?,\ \globalState'[j])\ \in \ \localTransRel$ &   & (7, $\land$) \\
	10. &  & $\forall k' \notin \{i, j\} \neq i': \altGlobalState[k']\ =\ \globalState[k']$ &   & (7, $\land$) \\
	11. &  & $\forall \globalPerm, \boundedEvent: (\globalPerm(\globalState), \boundedEvent, \boundedAltGlobalState) \notin \bounded{\globalTransRel}$ &   & (4, $\neg \exists$) \\
	12. &  & $\forall \globalPerm:(\globalPerm(\globalState), \globalPerm'[j](\event), \boundedAltGlobalState) \notin \bounded{\globalTransRel}$ &   & (10, $\forall \boundedEvent = \globalPerm'[j](\event)$) \\
	13. &  & $\globalPerm = \mkperm(\globalPerm', \globalState, \altGlobalState) \ \land\ (\globalPerm(\globalState), \globalPerm'[j](\event), \boundedAltGlobalState) \notin \bounded{\globalTransRel}$ &   & (12, $\forall \globalPerm = \mkperm(\globalPerm', \globalState, \altGlobalState)$) \\
	14. &  & $\globalPerm = \mkperm(\globalPerm', \globalState, \altGlobalState)$ &   & (12, $\land$) \\
	15. &  & $(\globalPerm(\globalState), \globalPerm'[j](\event), \boundedAltGlobalState) \notin \bounded{\globalTransRel}$ &   & (12, $\land$) \\
	16. &  & $\neg \exists i': (\globalPerm(\globalState)[i'],\ \globalPerm'[j](\event)!,\ \boundedAltGlobalState[i'])\ \in \ \localTransRel \lor$ &   &  \\
	&  & $\neg \exists j': (\globalPerm(\globalState)[j'],\ \globalPerm'[j](\event)?,\ \boundedAltGlobalState[j'])\ \in \ \localTransRel \lor $ &   &  \\
	&  & $\neg \forall k' \notin \{i', j'\}: \boundedAltGlobalState[k']\ =\ \globalPerm(\globalState)[k'] \lor $ &   &  \\
	&  & $\globalPerm(\globalState) \notin \bounded{\globalTransRel} \lor \boundedAltGlobalState \notin \bounded{\globalTransRel}$ &   & (15, pw-global) \\
	17. &  & $\forall i': (\globalPerm(\globalState)[i'],\ \globalPerm'[j](\event)!,\ \boundedAltGlobalState[i'])\ \notin \ \localTransRel \lor$ &   & \\
	&  & $\forall j': (\globalPerm(\globalState)[j'],\ \globalPerm'[j](\event)?,\ \boundedAltGlobalState[j'])\ \notin \ \localTransRel \lor$ &   & \\
	&  & $\exists k' \notin \{i', j'\}: \boundedAltGlobalState[k'] \neq \globalPerm(\globalState)[k'] \lor$ &   &  \\
	&  & $\globalPerm(\globalState) \notin \bounded{\globalTransRel} \lor \boundedAltGlobalState \notin \bounded{\globalTransRel}$ &   & (16, $\neg \exists$, $\neg \forall$) \\
	18. & a) & $\boundedAltGlobalState \notin \bounded{\globalTransRel}$ &   & (17, $\lor$) \\
	19. & a) & $\bot$ &   & (2, 18a) \\20. & b) & $\globalPerm(\globalState) \notin \bounded{\globalTransRel}$ &   & (17, $\lor$) \\
	21. & b) & $\globalPerm(\globalState) \in \bounded{\globalTransRel}$ &   & (2, bounded-upshot) \\
	22. & b) & $\bot$ &   & (20b, 21) \\
	23. &  & $\forall i': (\globalPerm(\globalState)[i'],\ \globalPerm'[j](\event)!,\ \boundedAltGlobalState[i'])\ \notin \ \localTransRel \lor$ &   &  \\
	&  & $\forall j': (\globalPerm(\globalState)[j'],\ \globalPerm'[j](\event)?,\ \boundedAltGlobalState[j'])\ \notin \ \localTransRel \lor $ &   &  \\
	&  & $\exists k' \notin \{i', j'\}: \boundedAltGlobalState[k'] \neq \globalPerm(\globalState)[k']$ &   & (17, 18-19a, 20-22b) \\
	24. &  & $(\globalPerm(\globalState)[i],\ \globalPerm'[j](\event)!,\ \boundedAltGlobalState[i])\ \notin \ \localTransRel \lor$ &   &  \\
	&  & $(\globalPerm(\globalState)[j],\ \globalPerm'[j](\event)?,\ \boundedAltGlobalState[j])\ \notin \ \localTransRel \lor$ &   &  \\
	&  & $\exists k' \notin \{i, j\}: \boundedAltGlobalState[k'] \neq \globalPerm(\globalState)[k']$ &   & (23, $\forall i' = i, j' = j$ (from 7)) \\
	25. &  & $(\globalPerm(\globalState)[i],\ \globalPerm'[j](\event)!,\ \boundedAltGlobalState[i])\ \notin \ \localTransRel \lor$ &   &  \\
	&  & $(\globalPerm(\globalState)[j],\ \globalPerm'[j](\event)?,\ \boundedAltGlobalState[j])\ \notin \ \localTransRel \lor$ &   &  \\
	&  & $k \notin \{i, j\}\ \land\  \boundedAltGlobalState[k] \neq \globalPerm(\globalState)[k]$ &   & (24, $\exists k' = k$ (arbitrary)) \\
	26. & a) & $k \notin \{i, j\}\ \land\  \boundedAltGlobalState[k] \neq \globalPerm(\globalState)[k]$ &   & (25, $\lor$) \\
	27. & a) & $k \notin \{i, j\}$ &   & (26, $\land$) \\
	28. & a) & $\boundedAltGlobalState[k] \neq \globalPerm(\globalState)[k]$ &   & (26, $\land$) \\
	29. & a) & $\globalPerm'(\altGlobalState)[k] \neq \globalPerm(\globalState)[k]$ &   & (28, 2) \\
	30. & a) & $\forall \varName \in \vars: k' \notin \{i, j\}:$ &   &  \\
	&    & $\getsFrom(\altGlobalState[k'], \varName, \globalState[k'], \varName)$ &   & (10, pw-global) \\
	31. & a) & $\forall \varName \in \vars: \getsFrom(\altGlobalState[k], \varName, \globalState[k], \varName)$ &   & (30, $\forall k' = k$ (from 25)) \\
	32. & a) & $\forall \varName' \in \vars: \globalPerm[j](\globalState[j](\varName)) = \globalPerm'[j](\globalState'[j](\varName))$ &   & (31, $\globalPerm$-kept) \\
	33. & a) & $\forall \varName' \in \vars: \globalPerm(\globalState)[j](\varName) = \globalPerm'(\altGlobalState)[j](\varName)$ &   & (32, $\globalPerm(\globalState)[i] = \globalPerm[i](\globalState[i])$) \\
	34. & a) & $\globalPerm(\globalState)[j] = \globalPerm'(\altGlobalState)[j]$ &   & (33, all vars equal) \\
	35. & a) & $\bot$ &   & (29a, 34a) \\
	36. & b) & $(\globalPerm(\globalState)[i],\ \globalPerm'[j](\event)!,\ \boundedAltGlobalState[i])\ \notin \ \localTransRel$ &   & (25, $\lor$) \\
	37. & b) & $(\globalPerm(\globalState)[i],\ \globalPerm'[j](\event)!,\ \globalPerm'(\altGlobalState)[i])\ \notin \ \localTransRel$ &   & (36b, 2) \\
	38. & b) & $(\globalPerm[i](\globalState[i]),\ \globalPerm'[j](\event)!,\ \globalPerm'[i](\globalState'[i]))\ \notin \ \localTransRel$ &   & (37b, $\globalPerm(\globalState)[i] = \globalPerm[i](\globalState[i])$) \\
	39. & b) & $(\globalPerm[i](\globalState[i]),\ \globalPerm[i](\event)!,\ \globalPerm[i](\globalState'[i]))\ \in \ \localTransRel$ &   & (8, $\localPerm$ preserves $\localTrans$) \\
	40. & b) & $\getsFrom(\globalState'[j], var_{e?}, \globalState[i], var_{e!})$ &   & (7, $\getsFrom$, pw-global) \\
	
\end{tabular}

\begin{tabular}{r r l c r}

41. & b) & $\globalPerm[i](\globalState[i](var_{e!})) = \globalPerm'[j](\globalState'[j](var_{e?}))$ &   & (40b, $\globalPerm$-gets) \\
	42. & b) & $\globalState[i](var_{e!}) = \delta$ &   & (8, pw-local) \\
	43. & b) & $\globalState'[j](var_{e?}) = \delta$ &   & (9, pw-local) \\
	44. & b) & $\globalPerm[i](\delta) = \globalPerm'[j](\delta)$ &   & (41b, 42b, 43b) \\
	45. & b) & $\globalPerm[i](\event) = \globalPerm'[j](\event)$ &   & (5, 44b) \\	46. & b) & $(\globalPerm[i](\globalState[i]),\ \globalPerm'[j](\event)!,\ \globalPerm[i](\globalState'[i]))\ \in \ \localTransRel$ &   & (39b, 45b) \\
	47. & b) & $\forall \varName' \in \vars \setminus havoc(\globalState'[i], \varName'): \exists \varName \in \vars:$ &   &  \\
	&    & $\getsFrom(\globalState'[i], \varName', \globalState[i], \varName)$ &   & (8, pw-internal-send) \\
	48. & b) & $\forall \varName' \in \vars \setminus havoc(\globalState'[i], \varName'):$ &   &  \\
	&    & $\globalPerm[i](\globalState[i](\varName')) = \globalPerm'[i](\globalState'[i](\varName))$ &   & (47b, $\globalPerm$-kept) \\
	49. & b) & $\forall \delta \in \Delta: \exists \varName \in \vars \setminus havoc(\globalState'[i], \varName) \land$ &   &  \\
	&    & $\delta = \globalState[i](\varName) \implies \globalPerm[i](\delta) = \globalPerm'[i](\delta)$ &   & (48b, variable eval) \\
	50. & b) & $(\globalPerm[i](\globalState[i]),\ \globalPerm'[j](\event)!,\ \globalPerm'[i](\globalState'[i]))\ \in \ \localTransRel$ &   & (46b, 49b, havoc-equiv) \\
	51. & b) & $\bot$ &   & (38b, 50b) \\
	52. & c) & $(\globalPerm(\globalState)[j],\ \globalPerm'[j](\event)?,\ \boundedAltGlobalState[j])\ \notin \ \localTransRel$ &   & (25, $\lor$) \\
	53. & c) & $(\globalPerm(\globalState)[j],\ \globalPerm'[j](\event)?,\ \globalPerm'(\altGlobalState)[j])\ \notin \ \localTransRel$ &   & (52c, 2) \\
	54. & c) & $(\globalPerm[j](\globalState[j]),\ \globalPerm'[j](\event)?,\ \globalPerm'[j](\globalState'[j]))\ \notin \ \localTransRel$ &   & (53c, $\globalPerm(\globalState)[i] = \globalPerm[i](\globalState[i])$) \\
	55. & c) & $(\globalPerm'[j](\globalState[j]),\ \globalPerm'[j](\event)?,\ \globalPerm'[j](\globalState'[j]))\ \in \ \localTransRel$ &   & (9, $\localTransRel$ preserves $T$) \\
	56. & c) & $\forall \varName, \varName' \in \vars :$ &   &  \\
	&    & $\getsFrom(\globalState'[j], \varName', \globalState[j], \varName) \implies$ &   &  \\
	&    & $ \globalPerm[j](\globalState[j](\varName)) = \globalPerm'[j](\globalState[j](\varName))$ &   & (14, $\globalPerm$-gets) \\
	57. & c) & $(\globalPerm[j](\globalState[j]),\ \globalPerm'[j](\event)?,\ \globalPerm'[j](\globalState'[j]))\ \in \ \localTransRel$ &   & (55c, 56c, src-equiv) \\
	58. & c) & $\bot$ &   & (54c, 57c) \\
	
\end{tabular}

All branches closed.   
	\end{footnotesize}

The proof is almost identical to the above when $\event$ is a broadcast transmission between system processes as well so, again, we elide the proof details for those transitions here.

Finally, we consider the case in which $\event$ is a consensus event (amongst non-environment processes). Again, we prove the lemma by contradiction, as follows.
	
	\begin{footnotesize}
		\begin{tabular}{r r l c r}
	
	1. &  & $(\globalState, \event, \altGlobalState) \in \globalTransRel$ &   & (assumption) \\
	2. &  & $\globalPerm'(\altGlobalState) = \boundedAltGlobalState \in \bounded{\globalStateSet}$ &   & (assumption) \\
	3. &  & $\altGlobalState \cwpRel \boundedAltGlobalState$ &   & (assumption) \\
	4. &  & $\neg\exists \globalPerm, \event':(\globalPerm(\globalState), \event', \boundedAltGlobalState) \in \bounded{\globalTransRel}$ &   & (assumption) \\
	5. &  & $\event = (a,\ \val^*,\ \text{vc},\ \bot)$ &   & (assumption) \\
	6. &  & $\exists S \subseteq I_n: \forall i \in S: \globalState[i](\text{ptct}_\event) = S$ &   & (assumption, participant set) \\
	7. &  & $\forall w' \in \val^*: \exists i' \in S: \globalState[i'](var_{\event!}) = w' \land$ &   &  \\
	&  & $\forall j' \in S: (\globalState[j'], \event?, \altGlobalState[j'])\ \in \ \localTransRel \land $ &   &  \\
	&  & $\forall k' \in I_n \setminus S: \altGlobalState[k'] = \globalState[k']$ &   & (1, vc-global) \\
	8. &  & $\forall w' \in \val^*: \exists i' \in S: \globalState[i'](var_{\event!}) = w'$ &   & (7, $\land$) \\
	9. &  & $\forall \globalPerm, \event': (\globalPerm(\globalState), \event', \boundedAltGlobalState) \notin \bounded{\globalTransRel}$ &   & (4, $\neg \exists$) \\
	10. &  & $\globalPerm = \mkperm(\globalPerm', \globalState, \altGlobalState) \ \land\ \forall \event': (\globalPerm(\globalState), \event', \boundedAltGlobalState) \notin \bounded{\globalTransRel}$ &   & (9, $\forall \globalPerm = \mkperm(\globalPerm', \globalState, \altGlobalState)$) \\
	11. &  & $\globalPerm = \mkperm(\globalPerm', \globalState, \altGlobalState)$ &   & (10, $\land$) \\
	12. &  & $\forall \event': (\globalPerm(\globalState), \event', \boundedAltGlobalState) \notin \bounded{\globalTransRel}$ &   & (10, $\land$) \\
	13. &  & $\forall i' \in [1, n]: \globalState[i'](var_{\event!}) = \altGlobalState[i'](var_{\event!})$ &   & (1, vc-local) \\
	14. &  & $\forall w' \in \val^*, i' \in S: \localPerm(w') = \globalPerm'[i'](w') \land$ &   &  \\
	&  & $ (\globalPerm(\globalState), \localPerm(\event), \boundedAltGlobalState) \notin \bounded{\globalTransRel}$ &   & (12, $\forall \event' = \localPerm(\event)$, $\val^*$ $stable$) \\
	
\end{tabular}

\begin{tabular}{r r l c r}

15. &  & $\forall w' \in \val^*, i' \in S: \localPerm(w') = \globalPerm'[i'](w')$ &   & (14, $\land$) \\
	16. &  & $(\globalPerm(\globalState), \localPerm(\event), \boundedAltGlobalState) \notin \bounded{\globalTransRel}$ &   & (14, $\land$) \\
	17. &  & $\exists w' \in \localPerm(\val^*): \forall i' \in S: \globalPerm(\globalState)[i'](var_{\event!}) \neq w'\lor $ &   &  \\
	&  & $\exists w' \in \localPerm(\val^*): \forall i' \in S: \globalPerm(\globalState)[i'](var_{\event!}) \neq w'\lor $ &   &  \\
	&  & $\exists j' \in S: (\globalPerm(\globalState)[j'], \localPerm(\event)?, \boundedAltGlobalState[j'])\ \notin \ \localTransRel \lor$ &   &  \\
	&  & $\exists k' \in I_n \setminus S: \boundedAltGlobalState[k'] \neq \globalPerm(\globalState)[k'] \lor$ &   &  \\
	&  & $ \globalPerm(\globalState) \notin \bounded{\globalStateSet} \lor \boundedAltGlobalState \notin \bounded{\globalStateSet}$ &   & (16, vc-global) \\
	18. & a) & $\boundedAltGlobalState \notin \bounded{\globalStateSet}$ &   & (17, $\lor$) \\
	19. & a) & $\bot$ &   & (2, 18a) \\
	20. & b) & $\globalPerm(\globalState) \notin \bounded{\globalStateSet}$ &   & (17, $\lor$) \\
	21. & b) & $\globalPerm(\globalState) \in \bounded{\globalStateSet}$ &   & (2, bounded-upshot) \\
	22. & b) & $\bot$ &   & (20a, 21a) \\
	23. & c) & $\exists k' \in I_n \setminus S: \boundedAltGlobalState[k'] \neq \globalPerm(\globalState)[k']$ &   & (17, $\lor$) \\
	24. & c) & $k \in I_n \setminus S \land \boundedAltGlobalState[k] \neq \globalPerm(\globalState)[k]$ &   & (23, $\exists k' = k$(arbitrary)) \\
	25. & c) & $k \in I_n \setminus S$ &   & (24, $\land$) \\
	26. & c) & $\boundedAltGlobalState[k] \neq \globalPerm(\globalState)[k]$ &   & (24, $\land$) \\
	27. & c) & $\globalPerm'(\altGlobalState)[k] \neq \globalPerm(\globalState)[k]$ &   & (28, 2) \\
	28. & c) & $\forall k' \in I_n \setminus S: \altGlobalState[k'] = \globalState[k']$ &   & (7, $\land$) \\
	29. & c) & $\forall \varName \in \vars, k' \in I_n \setminus S:$ &   &  \\
	&    & $\getsFrom(\altGlobalState[k'], \varName, \globalState[k'], \varName)$ &   & (28, vc-global) \\
	30. & c) & $\forall \varName \in \vars: \getsFrom(\altGlobalState[k], \varName, \globalState[k], \varName)$ &   & (29c, $\forall k' = k$ (from 24)) \\
	31. & c) & $\forall \varName' \in \vars: \globalPerm[k](\globalState[k](\varName)) = \globalPerm'[k](\altGlobalState[k](\varName))$ &   & (30c, $\globalPerm$-gets) \\
	32. & c) & $\forall \varName' \in \vars: \globalPerm(\globalState)[k](\varName) = \globalPerm'(\altGlobalState)[k](\varName)$ &   & (31c, $\globalPerm(\globalState)[i] = \globalPerm[i](\globalState[i])$) \\
	33. & c) & $\globalPerm(\globalState)[k] = \globalPerm'(\altGlobalState)[k]$ &   & (32c, all vars equal) \\
	34. & c) & $\bot$ &   & (27c, 33c) \\
	35. & d) & $\exists j' \in S: (\globalPerm(\globalState)[j'], \localPerm(\event)?, \boundedAltGlobalState[j'])\ \notin \ \localTransRel$ &   & (17, $\lor$) \\
	36. & d) & $j \in S \land (\globalPerm(\globalState)[j], \localPerm(\event)?, \boundedAltGlobalState[j])\ \notin \ \localTransRel$ &   & (35d, $\exists j' = j$(arbitrary))) \\
	37. & d) & $j \in S$ &   & (36d, $\land$) \\
	38. & d) & $(\globalPerm(\globalState)[j], \localPerm(\event)?, \boundedAltGlobalState[j])\ \notin \ \localTransRel$ &   & (36d, $\land$) \\
	39. & d) & $(\globalPerm(\globalState)[j], \localPerm(\event)?, \globalPerm'(\altGlobalState)[j])\ \notin \ \localTransRel$ &   & (38d, 2) \\
	40. & d) & $(\globalPerm[j](\globalState[j]), \localPerm(\event)?, \globalPerm'[j](\altGlobalState[j]))\ \notin \ \localTransRel$ &   & (39d, $\globalPerm(\globalState)[i] = \globalPerm[i](\globalState[i])$) \\
	41. & d) & $\forall j' \in S: (\globalState[j'], \event?, \altGlobalState[j'])\ \in \ \localTransRel$ &   & (7, $\land$) \\
	42. & d) & $(\globalState[j], \event?, \altGlobalState[j])\ \in \ \localTransRel$ &   & (41d, 37d, $\forall j' = j$ (from 36d)) \\
	43. & d) & $(\globalPerm'[j](\globalState)[j], \globalPerm'[j](\event)?, \globalPerm'[j](\altGlobalState)[j])\ \in \ \localTransRel$ &   & (42d, $\localPerm$ preserves $T$) \\
	44. & d) & $\forall \varName, \varName' \in \vars :$ &   &  \\
	&    & $\getsFrom(\altGlobalState[j], \varName', \globalState[j], \varName) \implies$ &   &  \\
	&    & $\globalPerm[j](\globalState[j](\varName)) = \globalPerm'[j](\globalState[j](\varName))$ &   & (11, $\globalPerm$-gets) \\
	45. & d) & $(\globalPerm[j](\globalState)[j], \globalPerm'[j](\event)?, \globalPerm'[j](\altGlobalState)[j])\ \in \ \localTransRel$ &   & (43d, 44d, src-equiv) \\
	46. & d) & $\forall w' \in \val^*: \localPerm(w') = \globalPerm'[j](w')$ &   & (15, 37d) \\
	47. & d) & $\localPerm(\event) = \globalPerm'[j](\event)$ &   & (46d, all vals equal) \\
	48. & d) & $(\globalPerm[j](\globalState)[j], \localPerm(\event)?, \globalPerm'[j](\altGlobalState)[j])\ \in \ \localTransRel$ &   & (45d, 47d, src-equiv) \\
	49. & d) & $\bot$ &   & (40d, 48d) \\
	50. & e) & $\exists w' \in \localPerm(\val^*): \forall i' \in S: \globalPerm(\globalState)[i'](var_{\event!}) \neq w'$ &   & (17, $\lor$) \\
	51. & e) & $w \in \localPerm(\val^*) \land \forall i' \in S: \globalPerm(\globalState)[i'](var_{\event!}) \neq w$ &   & (50e, $\exists w' = w$(arbitrary)) \\
	52. & e) & $w \in \localPerm(\val^*)$ &   & (51e, $\land$) \\
	53. & e) & $\forall i' \in S: \globalPerm(\globalState)[i'](var_{\event!}) \neq w$ &   & (51e, $\land$) \\
	54. & e) & $\forall w' \in \val^*: \exists i' \in S: \globalState[i'](var_{\event!}) = w'$ &   & (7, $\land$) \\
	55. & e) & $\exists i' \in S: \globalState[i'](var_{\event!}) = \localPerm^{-1}(w)$ &   & (52e, 54e) \\
	
\end{tabular}

\begin{tabular}{r r l c r}

56. & e) & $i \in S \land \globalState[i](var_{\event!}) = \localPerm^{-1}(w)$ &   & (55e, $\exists i' = i$(arbitrary)) \\
	57. & e) & $i \in S$ &   & (56e, $\land$) \\
	58. & e) & $\globalState[i](var_{\event!}) = \localPerm^{-1}(w)$ &   & (56e, $\land$) \\
	59. & e) & $\globalPerm(\globalState)[i](var_{\event!}) \neq w$ &   & (53e, 57e, $\forall i' = i$ (from 36d)) \\
	60. & e) & $\globalPerm'[i](\globalState[i](var_{\event!})) = w$ &   & (57e, 58e, 15) \\
	61. & e) & $\getsFrom(\altGlobalState[i], var_{\event!}, \globalState[i], var_{\event!})$ &   & (13, vc-local) \\
	62. & e) & $\globalPerm[i](\globalState[i](var_{\event!})) = \globalPerm'[i](\globalState[i](var_{\event!}))$ &   & (61e, 37d, $\globalPerm$-gets) \\
	63. & e) & $\globalPerm[i](\globalState[i](var_{\event!})) = w$ &   & (62e, 60e) \\
	64. & e) & $\globalPerm[i](\globalState[i](var_{\event!})) \neq w$ &   & (59e, $\globalPerm(\globalState)[i] = \globalPerm[i](\globalState[i])$) \\
	65. & e) & $\bot$ &   & (63e, 64e) \\
	
\end{tabular}

All branches closed.   
	\end{footnotesize}
	
	Having exhausted all types of events that $\event$ can be, we can finally conclude that it is indeed the case that, using our construction of $\globalPerm = \mkperm(\globalPerm', \globalState, \altGlobalState)$, it is always possible to identify an event $\boundedEvent$ such that $\exists (\boundedGlobalState, \boundedEvent, \boundedAltGlobalState) \in \bounded{\globalTransRel}:
	\globalState \cwpRel \boundedGlobalState$.
\end{proof}





%
%
%
%
%
%

\end{document}